\documentclass[prd, aps, 10pt, twocolumn, nofootinbib, superscriptaddress, preprintnumbers, balancelastpage, longbibliography, floatfix]{revtex4-2}
\usepackage[margin=1in]{geometry}
\usepackage{amsmath,amssymb,mathtools, physics, bm, soul, graphicx, xcolor, booktabs}
\usepackage[english]{babel}
\usepackage{comment}

\usepackage[breaklinks=true]{hyperref}

\hypersetup{
     colorlinks = true,
     citecolor  = blue,
     urlcolor   = purple,
     linkcolor  = magenta
}

\newcommand{\Tfo}{T_{\rm fo}}

\newcommand{\TQCD}{T_{\rm QCD}}
\newcommand{\LQCD}{\Lambda_{\rm QCD}}
\newcommand{\TBBN}{T_{\rm BBN}}

\newcommand{\Bs}{B_{\rm strange}}
\newcommand{\Pcoal}{P_{\rm coal}}

\newcommand{\mm}[1]{\textbf{ {[MM: #1]}}}

\begin{document}

\preprint{MIT--CTP/5975}

\title{Non-Thermal Production of Sexaquark Dark Matter}
\author{Marianne Moore}
\email{mamoore@mit.edu}
\affiliation{Center for Theoretical Physics -- a Leinweber Institute, Massachusetts Institute of Technology, Cambridge, MA 02139, USA}

\author{Stefano Profumo}
\email{profumo@ucsc.edu}
\affiliation{Department of Physics, 
University of California Santa Cruz, Santa Cruz, CA 95064, USA}
\affiliation{Santa Cruz Institute for Particle Physics,
Santa Cruz, CA 95064, USA}

\date{\today}

\begin{abstract}

Standard thermal freeze-out scenarios with QCD-scale interaction rates predict a $uuddss$ sexaquark relic abundance many orders of magnitude below the observed dark matter density, representing a key challenge for sexaquark dark matter models. Additionally, if the maximum post-inflationary temperature never exceeds the QCD confinement scale, the usual thermal/chemical-equilibrium production of the sexaquark near ${T \sim T_{\rm QCD} \simeq 150 \text{-} 170~\mathrm{MeV}}$ never occurs. In this work we show that non-thermal mechanisms can naturally overcome this obstacle. Using late-decaying reheatons as a representative case (while noting the broader applicability), we demonstrate that the final abundance is determined by two quantities: the branching fraction into strange-quark-rich matter and the coalescence probability into sexaquarks during the matter-dominated or early radiation-dominated epoch. We provide compact expressions and benchmark calculations for reheating temperatures $T_R \in [10, 100]~\mathrm{MeV}$ and reheaton masses above the QCD confinement scale. 
Unlike the predictive but unsuccessful thermal scenario, non-thermal production is sensitive to injection microphysics, coalescence efficiency, and residual entropy dilution. We delineate the viable parameter space, evaluate collider and precision constraints on representative reheaton models, and derive indirect detection bounds on residual antisexaquark populations. Our results establish non-thermal production as a viable pathway to sexaquark dark matter and highlight broader implications for non-equilibrium mechanisms in the early universe.

\end{abstract}

\maketitle

\section{Introduction}

The sexaquark, hereafter denoted by $S$, is a hypothetical composite dark matter candidate made of six Standard Model~(SM) quarks: two up quarks, two down quarks, and two strange quarks ($uuddss$)~\cite{Farrar:2017eqq}. 
It is electrically neutral, spinless, and, if sufficiently compact, potentially cosmologically stable.
The most plausible mass window for stability corresponds to 1860 to 1890~MeV, near the mass of two nucleons~\cite{Kolb:2018bxv}.
Since the sexaquark does not require beyond-the-Standard-Model interactions or particles, it offers in principle an appealing dark matter (DM) candidate. 
Its viability hinges on an unusually small radius, {$r_S \sim 0.1$-$0.4$~fm}, consistent with the absence of a surrounding pion cloud~\cite{Farrar:2017eqq}. 
Such compactness would suppress interactions with nucleons, preventing binding to nuclei, avoiding large scattering rates that would be visible in dark matter direct detection experiments, and making production at colliders far smaller than typical baryon production rates.

In addition to these attractive features, a number of challenges for the sexaquark scenario have been noted.
Reproducing the correct cosmological abundance is nontrivial: early-universe production must be efficient enough despite the strong suppression of multi-quark bound-state formation in thermal QCD environments~\cite{Farrar:2020zeo, Blaschke:2021tul}.
On the experimental side, the lack of observed anomalously stable multi-strange nuclei, null results from double-hypernuclei searches, and bounds from nuclear stability all constrain the allowed properties of such a state~\cite{doi:10.1098/rspa.1989.0115, Aoki:1991ip, Gal:2016boi}.

A central challenge for sexaquark dark matter arises in the cosmological scenario where the Universe thermalizes above the QCD crossover and sexaquarks subsequently freeze out in a hadronic bath.
Because QCD-scale interactions keep sexaquarks in chemical equilibrium with baryons and photons far below the QCD transition, efficient breakup and annihilation channels continue to deplete their abundance until freeze out occurs, producing an exceedingly small yield. 
The thermal relic density of sexaquarks thus falls significantly short of the observed dark matter abundance, by over ten orders of magnitude~\cite{Kolb:2018bxv, Moore:2024mot}. 
Unless all relevant sexaquark interaction rates are suppressed by many orders of magnitude compared to the natural QCD scale, the thermal production mechanism cannot therefore account for the observed dark matter density.

Non-thermal production mechanisms can naturally circumvent this problem by generating sexaquarks after the efficient equilibrium annihilation era has ended, or through processes that are never in equilibrium with the thermal bath. 
In particular, since cosmology prior to Big Bang Nucleosynthesis~(BBN) is only weakly, if at all, constrained, it is possible for reheating to complete below the QCD crossover, but still safely above temperatures required for successful BBN, i.e. ${\TQCD \gg T_R \gg \TBBN}$.
As long as reheating completes before ${T \sim 5}~\text{MeV}$, this alternate cosmological scenario would not affect BBN observables~\cite{Kawasaki:2000en, Hannestad:2004px, deSalas:2015glj, Barbieri:2025moq}.

In this work, we explore non-thermal production mechanisms that naturally overcome the thermal abundance deficit. 
We focus on a concrete and calculable example: a low-reheating scenario where a late-decaying scalar field (the reheaton) decays predominantly into light SM degrees of freedom, which subsequently form sexaquarks during the hadronization and coalescence of a strange-rich plasma.
While we use the reheaton as an explicit realization and benchmark, the framework applies equally to the decay of any long-lived field in the same cosmological epoch, provided it injects hadronic matter below $\TQCD$ but above BBN temperatures.

We factorize the production of sexaquarks from decays into two major contributions: (i)~the prompt production, from the decays, into strange-quark-rich matter, and (ii)~the coalescence probability of such strange matter into sexaquarks. 
Together with the reheaton mass, couplings to SM states, and the decay temperature, these quantities fully control the resulting relic abundance of sexaquarks.
Using this framework, we determine the regions in parameter space where sexaquarks can be produced as all of dark matter. 
Moreover, we identify indirect detection constraints for such particles, as well as collider and precision frontier limits on the parent field when it is identified with a reheaton.

The paper is organized as follows. 
We begin by reviewing the thermal production of sexaquark dark matter in Sec.~\ref{sec:review}.
Then, we evaluate the required branching ratio of reheaton that will eventually generate sexaquarks and define our factorization of their production in Sec.~\ref{sec:factorize}. 
In Sec.~\ref{sec:Bstrange}, we evaluate the first term from factorization, the branching ratio into strange-rich matter, by exploring various reheaton models, the resulting strange-quark branching ratio, as well as the constraints on the reheaton models. 
Then, in Sec.~\ref{sec:coalescence}, we estimate the second term from factorization, the coalescence probability, and compare our results with previous works. 
We evaluate an alternate scenario, the potential sexaquark yield due to freeze-in during reheating due to the reheaton decay in Sec.~\ref{sec:freeze-in}. 
In Sec.~\ref{sec:antiparticles}, we investigate the depletion of antibaryons and antisexaquarks, baryogenesis, as well as indirect detection constraints of a non-negligible relic population of antisexaquarks.
We link the sexaquark-genesis to the baryon asymmetry, and introduce scenarios where the two can be simultaneously addressed in Sec.~\ref{sec:asymmetry}.
Finally, we summarize our findings in Sec.~\ref{sec:summary}.

Throughout this paper, we use ${g_\ast \simeq 10.75}$ as the effective number of light degrees of freedom in the MeV era~\cite{Kolb:1990vq} as we do not add any effective light degrees of freedom to the standard cosmological picture. 
We also use the observed baryon-to-photon ratio to be ${\eta \equiv n_b / n_\gamma \simeq 6 \times 10^{-10}}$, where ${n_b = \sum_B n_B - n_{\bar{B}}}$ is summed over baryon species $B$~\cite{PDG:2024}, which remains constant from the end of entropy injection (reheaton decay) to today's measured value.

We also take ${m_S = 1860}~\text{MeV}$ as the sexaquark mass.
Varying this mass in the range $[1860, 1890]$~MeV does not qualitatively affect our conclusions.
 {This mass range corresponds to the most plausible mass window for stability; a too heavy state could doubly-weakly decay to nucleons, while a too light state could be produced from the coalescence of two baryons in a nucleus~\cite{Kolb:2018bxv}. 
Estimates of the mass of this particle have been done using lattice QCD at unphysical pion masses, see e.g. Ref.~\cite{Green:2021qol} for a recent study and a summary of the results from the community. 
Additional techniques, namely the MIT quark bag model, QCD sum rules, diquark models, holographic models of QCD, find a bound $uuddss$ state with mass varying between 1200 and 2200~MeV~\cite{Jaffe1977, Kodama:1994np, Gross:2018ivp, Azizi:2019xla, Evans:2023zde}. 
While these studies indicate that a bound $uuddss$ state is plausible, its mass could plausibly lie outside the stability window considered here. 
The phenomenology of sexaquarks is extremely sensitive to the mass; only the narrow interval $[1860-1890]$~MeV permits a cosmologically stable relic compatible with existing experimental bounds.
The production mechanisms studied in this work therefore apply only if the true sexaquark mass lies within this window; otherwise, sexaquarks cannot constitute the dominant component of dark matter.}

\section{\label{sec:review}Review of thermal sexaquark production}

The standard cosmological scenario for the sexaquark assumes a high reheating temperature $T_R$ compared to the QCD crossover temperature $\TQCD$.
The sexaquark chemical freeze-out temperature $\Tfo$ lies below the crossover, giving ${T_R \gg \TQCD \gg \Tfo}$.
In this picture, the Universe first thermalizes, then evolves through a quark-gluon plasma, and subsequently hadronizes.
Sexaquarks freeze out in this thermal hadronic bath~\cite{Gross:2018ivp, Kolb:2018bxv, Moore:2024mot}.

However, this thermal scenario faces a severe relic abundance deficit.
The measured baryon asymmetry fixes the baryon-number chemical potential $\mu_B(T)$, 
and in chemical equilibrium  nucleons and sexaquarks carry chemical potentials 
${\mu_N = \mu_B}$ and ${\mu_S = 2\mu_B}$, respectively.
Since the sexaquark is heavier than a single nucleon, its nonrelativistic 
equilibrium abundance is Boltzmann suppressed relative to baryons.
Even if the sexaquark were slightly lighter than two nucleons, 
${m_S \lesssim 2 m_N}$, such that it is stable and can in principle be the lightest 
${B=2}$ state, its larger mass per particle leads to a suppressed equilibrium yield.

For temperatures somewhat below the QCD crossover, e.g.~${T \sim 100~\mathrm{MeV}}$, the equilibrium sexaquark-to-nucleon ratio is roughly
\begin{equation}
  \frac{n_S^{\rm eq}}{n_N^{\rm eq}} 
  \sim \exp\left(-\frac{m_S - m_N}{T}\right) 
  \sim \mathcal{O}(10^{-4}) \ ,
\end{equation}
where we have taken ${m_S - m_N \simeq 922~\mathrm{MeV}}$ (for ${m_S = 1860~\mathrm{MeV}}$ and ${m_N = 938~\mathrm{MeV}}$) and neglected the baryon chemical potential, which is a reasonable assumption at these temperatures~\cite{Moore:2024mot}.
This suppression is clearly seen in detailed equilibrium calculations, where sexaquarks carry only a small fraction of the cosmic baryon asymmetry at all temperatures below the QCD crossover.
While rapid baryon-number-changing reactions keep the sexaquark abundance close to its equilibrium value, as the Universe cools and expands these reactions eventually become inefficient relative to the Hubble rate, and the sexaquark number density freezes out at a value far below that required to account for the observed dark matter density when using QCD-scale interaction rates.

As emphasized in Refs.~\cite{Kolb:2018bxv, Moore:2024mot}, baryon-exchange reactions of the form ${S X \leftrightarrow B B'}$, with $B,B'$ octet baryons and $X$ light mesons or photons, remain rapid in the thermal bath after hadronization.
While the breakup direction ${S X \to B B'}$ is endothermic in vacuum, at temperatures ${T\sim 100 \text{-} 200~\mathrm{MeV}}$ thermal excitation makes it efficient.
These reactions enforce ${\mu_S = 2\mu_B}$ and drive the system toward its equilibrium configuration, in which almost all baryon number resides in nucleons and the sexaquark fraction is suppressed by several orders of magnitude. Consequently, efficient ${S X \to B B'}$ exchange leads to a \emph{depletion} of the sexaquark abundance rather than an enhancement.

For typical QCD-scale cross sections ${\sigma \sim 10~{\rm mb}}$, the annihilation rate ${\Gamma_{\rm ann} = n_S \langle \sigma v \rangle}$ remains larger than the Hubble rate down to very low temperatures, driving the sexaquark abundance to a tiny freeze-out value.
\citet{Kolb:2018bxv} showed that irrespective of the initial hadron abundances produced by the QCD transition, the final sexaquark number density is
\begin{equation}
    \frac{n_S}{n_B} \lesssim 10^{-11} \ ,
\end{equation}
which corresponds  to a thermal relic density of
\begin{equation}
    \frac{\Omega_S^{\rm thermal}}{\Omega_{\rm DM}} \lesssim 10^{-11} \ .
\end{equation}
Achieving the observed dark matter abundance, $\Omega_{\rm DM} h^2 \simeq 0.12$, would require suppressing the annihilation cross section by approximately seventeen orders of magnitude below typical QCD scales.
A number of works have suggested that part of this suppression (up to twelve orders of magnitude) might arise naturally if the sexaquark is extremely compact, yielding minimal spatial overlap with ordinary baryons and hyperons~\cite{Farrar:2017eqq, Farrar:2018hac, McDermott:2018ofd}. 
In this picture, transition amplitudes between sexaquarks and conventional hadrons would be strongly suppressed, simultaneously weakening nuclear binding effects, reducing decay and conversion rates in matter, and diminishing the sensitivity of many traditional experimental searches.

Such an extreme reduction would keep sexaquarks and antisexaquarks in near-symmetric abundances until their abundance freezes out with a near-identical yield, which is itself tightly constrained and ultimately inconsistent with the required dark matter density~\cite{Moore:2024mot}. 
Furthermore, no known mechanism has been claimed to suppress the interaction cross section by more than twelve orders of magnitude to obtain sexaquarks as a large fraction of dark matter.
This extreme fine tuning represents the fundamental obstacle to viable thermal sexaquark dark matter.

\section{\label{sec:factorize}Factorizing the sexaquark production}

In addition to the standard high-temperature reheating picture, it is well motivated to consider ``low reheating'' cosmologies in which the post-inflationary Universe remains matter dominated by a long-lived field $\phi$ until relatively late times. When this field (which we will generically call the \emph{reheaton}) decays at a temperature $T_R$ below the QCD crossover but still safely above the onset of BBN, ${T_{\rm QCD} \gtrsim T_R \gtrsim {\cal O}(5~{\rm MeV})}$, reheating and hadron production occur directly from reheaton decays rather than from a fully thermalized quark-gluon plasma. Such late-decay scenarios arise naturally for moduli, inflatons, and other weakly coupled scalars, and have been extensively studied in the context of BBN bounds and non-thermal particle production~\cite{Kawasaki:2000en, Giudice:2000ex, Allahverdi:2010xz}. In the following we adopt this framework and parametrize the cosmology in terms of the reheaton mass $m_\phi$, decay rate $\Gamma_\phi$ (corresponding to a lifetime ${\tau_\phi=\Gamma_\phi^{-1}}$), and the associated reheating temperature $T_R$.

In the reheaton decay scenario, by the end of reheating, the comoving number of decays per unit entropy is ${Y_\phi \simeq (3 T_R)/(4 m_\phi)}$, where $m_\phi$ is the reheaton mass, which does not need to be a scalar. 
Apart from $T_R$ and $m_\phi$, $Y_\phi$ depends only on energy conservation at the transition to radiation domination, and is therefore insensitive to the detailed microphysics of $\phi$ decay.
If each reheaton decay produces a sexaquark with probability $f_S$, then the resulting sexaquark yield is 
\begin{align}\label{eq:yield_sexaquark}
    Y_S^{\rm (dec)} \simeq f_S \times Y_\phi = f_S\frac{3T_R}{4m_\phi} \ .
\end{align}
Matching the present relic dark matter abundance, ${\Omega_{\rm DM}h^2\simeq 0.12}$, requires
\begin{align}
  f_S^{\rm req} \simeq 
  3.1 \times 10^{-7} 
  \left( \frac{m_\phi/T_R}{10^3} \right) 
  \left( \frac{1860~\mathrm{MeV}}{m_S} \right) \ ,
  \label{eq:fSreq}
\end{align}
which scales linearly with the ratio $m_\phi/T_R$: a higher reheating temperature or lighter reheaton can relax the needed efficiency proportionally. 
We note that a substantially identical yield would derive from the decay of a massive field at the same epoch, even if this field were unrelated to reheating.

\begin{figure}[ttt]
    \centering
    \includegraphics[width=\linewidth]{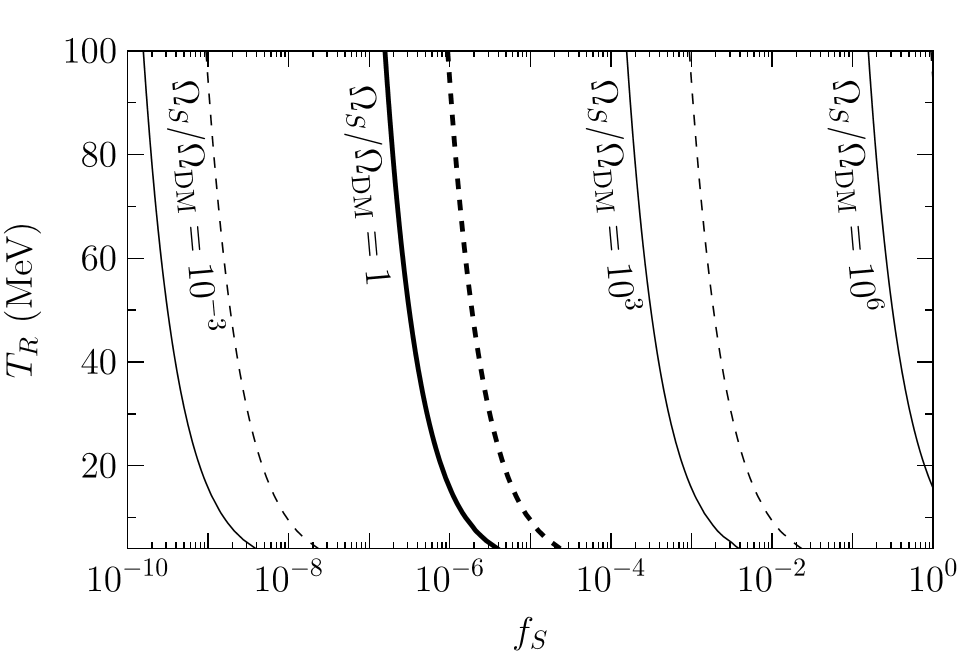}
    \caption{Dark matter abundance in sexaquarks, $\Omega_S / \Omega_\text{DM}$, as a function of the reheating temperature $T_R$ and the fraction of reheaton decays into sexaquarks $f_S$. The bold lines indicate the parameter combinations where sexaquarks constitute the entirety of dark matter (${\Omega_S = \Omega_\text{DM}}$) for two representative reheaton masses: ${m_\phi = 50~\text{GeV}}$~(solid) and ${m_\phi = 300~\text{GeV}}$~(dashed).}
    \label{fig:omegaS}
\end{figure}

Throughout this work, we consider benchmark reheating temperatures ${T_R \in [10,100]~\text{MeV}}$, which lie between the QCD crossover temperature and the onset of BBN, and reheaton masses ${m_\phi \gg \LQCD}$, for which values of $f_S$ can be sufficient to reproduce the observed dark matter density.
Figure~\ref{fig:omegaS} presents this parameter space for two choices of the reheaton mass $m_\phi$.
Larger reheaton masses require a correspondingly larger fraction $f_S$ to produce the same abundance.
We note that in the figure, we neglect effects such as freeze-in production and relic annihilation, which we discuss below.
Furthermore, it is in principle possible to overproduce sexaquarks, and subsequently deplete the excess population via (self-)annihilation, provided this interaction rate is strong enough.
This would correspond to the top right corner of Fig.~\ref{fig:omegaS}, for large reheating temperatures and fractions $f_S$, although achieving a large $f_S$ is unlikely to be realized given the suppression factors inherent to coalescence.
Decaying at ${T \simeq T_R}$ with ${\TQCD \gg T_R \gg \TBBN}$ also implies that the reheaton lifetime is in the range ${\tau_\phi \in [10^{-5},1]}$~s, which will have consequences for detecting this particle in collider experiments.

Since we consider scenarios where the reheaton mass is much larger than the QCD scale, ${m_\phi \gg \LQCD}$, we can factorize the efficiency $f_S$ into a first factor, $\Bs$, that estimates partonic branchings into strange-rich matter, and a second factor, $\Pcoal$, that folds in the non-equilibrium probability that the strange-rich system coalesces into a compact, long-lived $S$.
The efficiency of sexaquark production from reheaton decay is then
\begin{align}\label{eq:fS_parametrization}
    f_S \equiv \Bs \times \Pcoal \ .
\end{align}
Both factors are highly model-dependent and subject to only order-of-magnitude constraints. 
In generic low-reheating models with no special strange enhancement, one may estimate the fraction of dark matter in sexaquarks, $\Omega_S/\Omega_\text{DM}$, to be generically below the observed dark matter density.
Reaching ${\Omega_S \simeq \Omega_{\rm DM}}$ thus likely requires tuned (but quantifiable) combinations of $\Bs$, $\Pcoal$, $m_\phi$, and $T_R$.
We explore in Sec.~\ref{sec:Bstrange} the magnitude of the first factor of Eq.~\eqref{eq:fS_parametrization}, $\Bs$, as a function of $m_\phi$.
We then investigate the second factor of Eq.~\eqref{eq:fS_parametrization}, $\Pcoal$, in Sec.~\ref{sec:coalescence}.

\section{\label{sec:Bstrange}Strange quark injection from reheaton decay}

\begin{figure}[ttt]
    \centering
    \includegraphics[width=\linewidth]{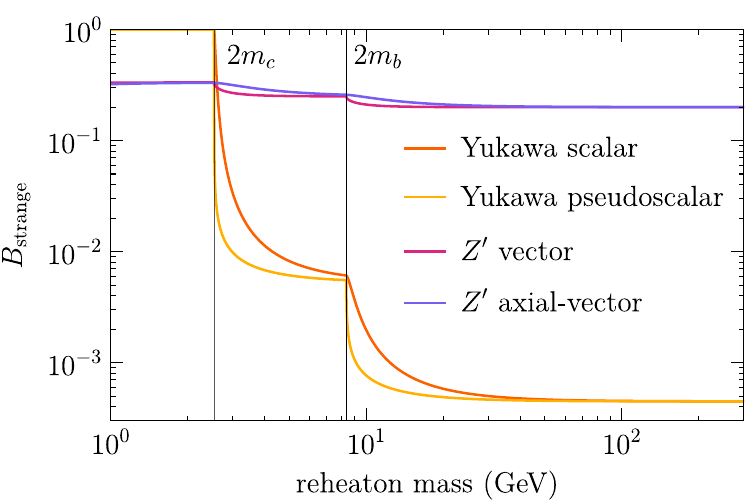}
    \caption{Branching ratios as a function of the reheaton mass $\phi$ from Eqs.~\eqref{eq:Bs_Yukawa} (Yukawa scalar in orange and pseudoscalar in yellow) and \eqref{eq:Bs_universal} (flavor-universal $Z'$ vector in magenta and axial-vector in violet). The thresholds for the reheaton to decay into two charm quarks~($2m_c$) and two bottom quarks~($2m_b$) are illustrated.}
    \label{fig:branching_ratios}
\end{figure}

In this section we map out the dominant mechanisms through which a late-decaying reheaton could inject strange quarks into the post-inflationary plasma.
Because the microscopic origin of the reheaton is model-dependent and different operator structures would populate strange quarks with very different efficiencies, we organize our analysis around several representative benchmark classes: scalar or reheatons with Yukawa-like (pseudo)scalar couplings, a gluon-coupled (``gluophilic'') scalar, and a (flavor-universal) vector or axial mediator.
Each scenario yields a characteristic strange-pair branching ratio $\Bs$ and distinct decay thresholds, illustrated in Fig.~\ref{fig:branching_ratios}.
Studying these different possibilities allows us to map out the range of feasible strange-quark injection histories and identify which regions of parameter space can realistically support efficient sexaquark production.

For each type of reheaton we separate the discussion into two parts.
First, we determine how the coupling structure controls the probability to produce strange quarks, including mass-threshold effects and the competition with other decay channels.
Second, we examine phenomenological constraints associated with creating and observing the reheaton itself.
This distinction is useful because the Universe may undergo a long matter-dominated reheating phase in which the reheaton lifetime can be as long as $\tau_\phi \sim 1$~s, allowing its decays to occur just before BBN.
In this regime, collider bounds can sometimes be parametrically weakened, particularly for particles with suppressed direct production.
However, we also comment on the separate case where the reheaton could be produced promptly at colliders even if a cosmological mechanism delays its decay to BBN scales; in that case, efficient production would imply that laboratory constraints cannot be evaded.
By presenting both the cosmological and experimental considerations together, we obtain a complete picture of how each reheaton types could populate the strange sector in a manner compatible with successful sexaquark production.

\subsection{\label{sec:Yukawa}Scalar or pseudoscalar reheaton with Yukawa-like couplings}

A scalar reheaton may couple to quarks via dimensionless Yukawa couplings $y_q$, which are proportional to the quark mass $m_q$ as ${y_q = m_q / \Lambda}$, where $\Lambda$ is a heavy mass scale.

The general Lagrangian quark-reheaton coupling term can be written as
\begin{align}\label{eq:pseudoscalar}
    \mathcal{L}_Y \supset -  \bar{q} (y_{qS} + i \gamma_5 y_{qP}) q \phi \ ,
\end{align}
where ${\beta_q=\sqrt{1-4m_q^2/m_\phi^2}}$ and $y_{qS}~(y_{qP})$ is the scalar~(pseudoscalar) Yukawa coupling. 
Analogous couplings to charged leptons $\ell$ can also be implemented with ${y_\ell = m_\ell / \Lambda}$.

\subsubsection{Strange quark injection from reheaton decay}

The decay rate of the reheaton into a pair of quarks $\phi \to q \bar{q}$ from this Lagrangian is
\begin{align}
    \Gamma_Y(\phi \to q \bar{q}) = \frac{N_c y_q^2 m_\phi}{8\pi} \left[ y_{qS}^2 \beta_q^3  + y_{qP}^2 \beta_q \right]
\end{align}
The branching ratio into strange quarks depends on which quark channels are opened, and is given by 
\begin{align}\label{eq:Bs_Yukawa}
    \Bs^Y = \frac{m_s^2 \left[ y_{qS}^2 \beta_s^3 + y_{qP}^2 \beta_s \right]}{\sum_{{\rm open}~q} m_q^2 \left[ y_{qS}^2 \beta_q^3  + y_{qP}^2 \beta_q \right]}
\end{align}
Away from quark mass thresholds, considering a pure scalar or pure pseudoscalar coupling gives similar numbers. 
For ${m_\phi > 2 m_b}$, the branching ratio is ${\Bs^Y \sim 10^{-3}}$, with smaller reheaton masses resulting in larger values of $\Bs$.
Opening reheaton decay channels to leptons would reduce the overall branching ratio into strange quarks.

Finally, we remark that enforcing the condition ${\Gamma(\phi \to q \bar{q}) \simeq H(T_R)}$, which ensures that reheating completes at temperature $T_R$ rather than significantly earlier or later, fixes the coupling strength of $\phi$ to quarks, and therefore determines the scale $\Lambda$. 
Solving this relation gives ${\Lambda \gtrsim 10^{11}}$~GeV. 
This requirement reflects the fact that the Universe transitions to radiation domination precisely when $\phi$ decays efficiently.

\subsubsection{Reheaton production pathways and observable signatures}

Scalar singlet extensions to the SM sector are strongly constrained by Higgs and electroweak precision searches at the LHC, LEP, as well as by fixed-target and flavor experiments~\cite{Hooper:2025iii}.
In practice, the allowed parameter space for a scalar or pseudoscalar reheaton is determined jointly by the limits on Higgs mixing (governing production in the large-mixing regime), bounds on exotic decays in heavy-flavor systems (governing production for ${m_\phi < m_B}$), and displaced vertex and long-lived particles (LLP) searches (relevant due to the cosmological scale $\Lambda$ implies extremely small Yukawa couplings, and thus macroscopic length scales).

The collider phenomenology of a CP-even scalar $\phi$ is governed by its mixing with the SM Higgs boson. 
Even if the primary coupling is to quarks via the scale $\Lambda$, radiative loop effects induce mixing.
The scalar mass matrix in the $\mqty(h & \phi)$ basis,
\begin{align}
    \mathcal{M}^2 = \mqty(m_h^2 & \delta m^2 \\ \delta m^2 & m_\phi^2) \ ,
\end{align}
is diagonalized by a mixing angle $\theta$, where ${\tan 2\theta = 2\delta m^2 / (m_h^2 - m_\phi^2)}$.
The off-diagonal term $\delta m^2$ arises from quark loops, and even a modest value could yield a large $\theta$.
In the limit where the masses are not degenerate, ${\delta m^2 \ll \abs*{m_h^2 - m_\phi^2}}$, the mixing effects remain negligible, ${\theta \simeq \delta m^2 / (m_h^2 - m_\phi^2)}$.
However, should the reheaton mass lie close to the Higgs mass, level repulsion would drive the system toward maximal mixing, causing $\phi$ to inherit Higgs-like production modes, such as gluon fusion and vector boson fusion, with rates scaled by $\sin^2\theta$.
In this regime, the reheaton's lifetime would become promptly short and would generically decay before $\TQCD$, thereby suppressing late-time hadronic injection.
Collider searches for exotic Higgs-width modifications or additional scalar resonances directly probe this mixing.

Although a tree-level Higgs-reheaton term ${\mathcal{L}\supset -\mu H^\dagger H \phi}$ could exist, it is already strongly constrained by LEP and LHC searches for exotic Higgs decays and additional scalar resonances~\cite{Robens:2015gla, Clarke:2013aya, Robens:2019kga}.
We therefore neglect this coupling and retain only the loop-induced mixing term $\delta m^2$, which provides a minimal and phenomenologically consistent portal between the reheaton and the SM sector.

If the reheaton contains a CP-odd admixture, the mass eigenstates inherit pseudoscalar couplings to SM fermions and gauge bosons. 
This induces two-loop Barr-Zee contributions to the electron, neutron, and atomic electric dipole moment~(EDM) involving a top quark or $W$ boson, even in the abscence of direct couplings to electrons.
Schematically, the electron EDM contribution takes the form~\cite{Barr:1990vd}
\begin{align}\label{eq:EDM}
    d_e &\simeq \frac{e \alpha}{8 \pi^3} \frac{m_e}{v^2} \sin\xi \Big[ \kappa_t F_t(m_t^2 / m_\phi^2) \nonumber\\
    &\qquad + \kappa_W F_W (m_W^2 / m_\phi^2) + \dots \Big] \ ,
\end{align}
where $\xi$ is the CP-odd mixing angle between $\phi$ and the SM Higgs in the mass basis and $F_{t,W}$ are dimensionless Barr-Zee loop functions, with ${F_{t,W} = \mathcal{O}(1)}$ in the reheaton mass range we consider. Finally, ${\kappa_{t,W} = \sin\theta + y_f v/m_f}$ denote effective $\phi$-fermion and $\phi$-$W$ couplings relative to the SM Higgs. 
The contribution from the reheaton's Yukawa couplings $y_f$ is negligible due to the large scale $\Lambda$.
The contributions to the electron EDM from other fermions (quarks or leptons) with ${m_f \ll m_\phi}$ are suppressed by $m_f^2/m_\phi^2$.
The tightest EDM limits are currently ${\abs*{d_e} < 0.041 \times 10^{-28}~e}$~cm from electrons in molecular ions and under the influence of a large intramolecular electric field~\cite{Roussy:2022cmp, PDG:2024}.
Current EDM limits constrain the effective CP-odd coupling $\kappa_{t,W}\sin\xi$ to be very small; for $m_\phi$ between a few and a few hundred GeV, values ${\sin\xi \times (\kappa_t + \kappa_W) \lesssim 5 \times 10^{-3}}$ are typically necessary in the absence of tuned cancellations~\cite{Pospelov:2005pr, Chiang:2015fta}.
For our low-reheat scenarios, maintaining a tiny CP-odd admixture ensures EDM limits are automatically satisfied while the cosmology and collider phenomenology are governed by the CP-even portal.

The reheaton could be produced in colliders primarily through gluon fusion (via a top loop), and, at lower masses, through associated production with heavy flavor.
For ${m_\phi < m_B}$, further production channels open in heavy-flavor environments, including rare $B$ {-meson} decays such as ${B \to K^{(\ast)}\phi}$ and radiative quarkonium transitions ${\Upsilon(nS) \to \gamma\phi}$.
These channels render $B$ factories and LHCb particularly sensitive to light, hadrophilic scalars and pseudoscalars.
The leading searches applicable to Yukawa-coupled $\phi$ include prompt and displaced dimuon searches, low-mass dijet and displaced-dijet analyses, and displaced hadronic-vertex searches~\cite{BaBar:2009hug, ATLAS:2024lowmassdijet, ATLAS:2024PRDlowmass, CMS:2025DataScoutingDijet, PDG:2024}.
The strongest limits generally arise in the dimuon channel, while purely hadronic or displaced-hadronic modes are significantly less constraining; in fact, existing data imposes no model-independent lower bound on $m_\phi$.
Instead, the sensitivity maps onto the coupling-lifetime plane, with Belle~II and LHCb offering substantial reach for LLP signatures~\cite{BaBar:2009hug, Belle2:prospects2024, LHCb:LLPpotential2024, PDG:2024}.

Because the scale $\Lambda$ implied by our cosmological setup is very large, the Yukawa couplings ${y_q = m_q / \Lambda}$ are tiny. 
Consequently, any $\phi$ produced at a collider would typically be highly boosted and decay well outside of a detector. 
This ensures compatibility with the long lifetime required for late reheating, and suppresses contributions to the $Z$ width from ${Z \to q \bar{q} \phi}$ (where ${m_\phi < m_Z}$) to negligible levels~\cite{Cesarotti:2024rbh}.

Finally, the Yukawa-like couplings ${y_q = m_q / \Lambda}$ do not alter quark masses unless $\phi$ acquires a vacuum expectation value $v_\phi$. 
Even in this scenario, the correction $y_q v_\phi$ is negligible unless $v_\phi$ approaches $\Lambda$, and therefore plays no role in the phenomenology discussed above.

\subsection{\label{sec:gluophilic}Gluon-coupled (``gluophilic'') scalar reheaton}

If the reheaton primarily decays into gluons, ${\phi \to gg}$, all strange quarks in the final state originate from the subsequent fragmentation of the gluon jets. 
The Lagrangian has the form
\begin{align}\label{eq:Lagrangian_phiGG}
    \mathcal{L}_G \supset \frac{c}{\Lambda} \phi G_{\mu\nu}^a G^{a\,\mu\nu} \ ,
\end{align}
where $c$ a dimensionless constant of order unity and $\Lambda$ is a heavy scale.

\subsubsection{Strange quark injection from reheaton decay}

The decay rate is 
\begin{align}
    \Gamma_G(\phi \to g g) = \frac{2}{\pi} \frac{c^2}{\Lambda^2} m_\phi^3 \ .
\end{align}
In the regime of interest, the strange content is governed by the strangeness suppression factor $\gamma_s$ of the Lund string model, which characterizes how likely a string break is to produce an $s\bar{s}$ pair relative to a light $u\bar{u}$ or $d\bar{d}$ pair.
In the semiclassical tunneling picture of string breaking, the production probability for heavier quark flavors decreases exponentially. This leads to
\begin{align}\label{eq:gamma_s}
    \gamma_s \sim \exp(-\pi \big[ m_s^2 - m_{u,d}^2 \big] \kappa) \ ,
\end{align}
where $\kappa \simeq 0.18~\text{GeV}^2$ is the effective string tension extracted from nonperturbative studies of the static quark potential~\cite{Andersson:1983ia, Sjostrand:2006za, Sjostrand:2014zea}. 
Lattice determinations of the static quark potential support this value of $\kappa$~\cite{Bali:2000gf,Necco:2001xg}.
Because the tunneling probability depends on the transverse mass, heavier flavors are disfavored, and phenomenological fits to collider data typically yield ${\gamma_s \approx 0.2 \text{-} 0.3}$.

To estimate the strange-quark yield in gluon fragmentation relevant for our scenario, we note that producing two strange quarks within the same correlation volume requires two independent string breaks to generate $s\bar{s}$ pairs.
This gives an approximate strange-budget factor
\begin{align}\label{eq:Bs_gluon}
    \Bs^G \sim \gamma_s^2 \approx 0.04 \text{-} 0.09 \ ,
\end{align}
with the spread reflecting uncertainties in the hadronization tuning. 
A similar number can be obtained by making use of the Dokshitzer-Gribov-Lipatov-Altarelli-Parisi~(DGLAP) equation for the splitting of gluons in terms of quarks, but for the low energies of reheaton decay, the string-breaking estimate is expected to better approach the true branching ratio.

Although the Lund string picture provides a widely used framework for hadronization and successfully reproduces a broad range of $e^+ e^-$ collider data, its extrapolation to hadronic collisions comes with important uncertainties~\cite{Altmann:2025afh}.
The model assumes the color-connected system fragments are isolated strings in vacuum, an approximation that works reasonably well in clean, high-energy leptonic collisions.
In hadronic environments, where multiple color exchanges and overlapping strings can occur, Pythia's default fragmentation model and the underlying Lund string model underproduce strange hadrons relative to LHC measurements~\cite{ALICE:2016fzo}.

\subsubsection{Reheaton production pathways and observable signatures}

A gluophilic reheaton could be produced at hadron colliders through gluon fusion, and would decay predominantly via ${\phi \to gg}$. For prompt decays, this yields a narrow dijet signature. 
Relevant constraints come from low-mass dijet trigger-level and scouting analyses from high-energy $pp$ collisions at ATLAS and CMS~\cite{ATLAS:2024lowmassdijet, ATLAS:2024PRDlowmass, CMS:2025DataScoutingDijet}. 
These exclude generic cross sections for masses ${m_\phi \sim 200 \text{-}650}$~GeV. 
For lighter masses or smaller couplings, constraints arise from LLP searches looking for displaced hadronic decays in the calorimeter or muon system~\cite{ATLAS:HCALLLP2024, PDG:2024}. 

The least constrained region corresponds to very small effective gluon couplings. In this limit, $\phi$ has a very low production rate and becomes long-lived. 
The decays would occur in sparse detector regions or outside the tracker, significantly weakening prompt dijet constraints.
Practically, masses of $\mathcal{O}(10~\text{GeV})$ and below remain weakly constrained if the decays are displaced and hadronic, as flavor factories have limited reach for pure gluon portals.

In the parameter space relevant for late-time reheating (${10^{-5}~\text{s} \lesssim \tau_\phi \lesssim 1}$~s), the reheaton is effectively stable on collider scales.
Because it would escape the detector, all resonance and displaced vertex bounds would vanish.
A potential surviving collider search would be monojets ($gg \to g\phi$, where $\phi$ is invisible).
However, given the small effective coupling $1/\Lambda$ required for this lifetime $\tau_\phi$, the production rate would too small to be observable above the SM background.

Similarly to the Yukawa-coupled reheaton from Sec.~\ref{sec:Yukawa}, if $\phi$ couples directly to gluons there can be a small induced mixing with the SM Higgs through its loop-induced coupling, dominantly from top loop.

\subsection{\label{sec:flavor}Flavor-universal vector or axial reheaton}

Similar to the Yukawa coupled reheaton from Sec.~\ref{sec:Yukawa}, we parametrize a flavor-universal quark-coupled vector or axial $Z'$ by the interaction
\begin{align}\label{eq:vector}
    \mathcal{L}_U \supset Z_\mu' \bar{q} \gamma^\mu (g_{qV} + g_{qA} \gamma^5) q \ ,
\end{align}
where $g_{qV}$, $g_{qA}$ are the vector and axial quark couplings, respectively. Universal couplings are those for which both ${g_{qV} = g_V}$ and ${g_{qA} = g_{A}}$ are respected.

\subsubsection{Strange quark injection from reheaton decay}

The decay rate of the reheaton into a pair of quarks from this Lagrangian is
\begin{align}\label{eq:Br_universal}
  \Gamma_U(Z' \to q\bar q) &= \frac{N_c m_{Z'}}{12\pi} \beta_q \Bigg[ g_{qV}^2 \left( 1 + \frac{2 m_q^2}{m_{Z'}^2} \right) \nonumber \\
  &\qquad+ g_{qA}^2 \beta_q^2 \Bigg] \ ,
\end{align}
giving for universal couplings ${g_{qV} = g_V}$ and ${g_{qA} = g_A}$ and for a reheaton mass far above the production threshold ${m_{Z'} \gg m_q}$ the branching ratio
\begin{align}\label{eq:Bs_universal}
    \Bs^U \simeq \frac{1}{N_f^{\rm open}} \ .
\end{align}
With ${m_{Z'} > 2 m_b}$, the branching ratio is ${\Bs^U \sim 0.2}$.
Non-universal charges could rescale this to $g_s^2/\sum g_q^2$, allowing strange-enhanced scenarios with large $\Bs$.

\subsubsection{Reheaton production pathways and observable signatures}

At hadron colliders, the dominant production channel is ${q \bar{q} \to Z'}$, followed by ${Z' \to j j}$, where $j$ represents a hadronic jet. 
For moderate-to-large couplings, this leads to prompt dijet resonances.
Current collider sensitivities are dominated by ATLAS and CMS dijet resonance searches, including low-mass trigger-level analyses~\cite{ATLAS:2024lowmassdijet, CMS:2025DataScoutingDijet, ATLAS:2024PRDlowmass}.
For $m_{Z'}$ below a few GeV, additional constraints arise from $B$-meson decay, quarkonium transition, and precision measurements at flavor factories.
There, LHCb, Belle, BaBar, and dedicated LLP searches provide the leading bounds~\cite{PDG:2024}.

A special case is a leptophobic and strange-philic $Z'$ having suppressed couplings to $u$ and $d$ quarks relative to $s$ quarks. 
This suppression reduces production at the LHC due to smaller strange PDF support, and weakens reinterpretations of dijet resonance searches~\cite{ATLAS:2024lowmassdijet, ATLAS:2024PRDlowmass, CMS:2025DataScoutingDijet}. 
Furthermore, small couplings necessary for such models would result in long lifetimes, shifting the signal into long-lived particle territory where experimental gaps remain.

For the lifetimes required by our cosmological setup, the $Z'$ would traverse the entire detector apparatus before decaying.
Therefore, constraints from prompt dijets and displaced vertices are unlikely to apply.
The phenomenology mimics that of an invisible vector.
The remaining relevant bounds are likely coming exclusively from missing energy searches.

A $Z'$ with vector coupling $g_{qV}$ shares key phenomenological features with a dark photon: quark loops induce kinetic mixing with the SM photon, and at higher order, with the $Z$ boson.
The axial contribution to this mixing is negligible because the axial current is anomalous and its loop-induced mixing is mass-suppressed.
However, because the couplings required for the reheaton to decay after the QCD crossover are small, the induced mixing is expected to be negligible, far below the levels currently probed by current dark photon searches~\cite{Abdullahi:2023tyk}.

\subsection{Increasing the strangeness budget}

The effective strangeness yield per reheaton decay, $\Bs$, is largely controlled by the available decay channels and hadronization dynamics.
As shown in Fig.~\ref{fig:branching_ratios}, once the reheaton mass exceeds the charm and bottom thresholds, decays into heavy flavors rapidly dominate, suppressing the relative strange-quark fraction.
Hence, in the simplest flavor-universal scenarios, $\Bs$ cannot exceed the values already displayed there.
To genuinely {\it enhance} the strangeness budget, one must therefore modify either the microscopic couplings or the hadronization environment.

A first possibility is to introduce flavor-nonuniversal couplings favoring strange over heavy quarks.
If the reheaton couples proportionally to Yukawa matrices but with suppressed charm and bottom terms, e.g. via hierarchical or loop-induced effective operators, then $\Bs$ can remain near unity even for ${m_\phi > 2m_c}$.
Alternatively, hadronization itself may amplify strange-quark production.
As mentioned above, the Lund string model parametrizes strangeness suppression by ${\gamma_s \simeq 0.2 \text{-} 0.3}$, yet collider data show that $\gamma_s$ can reach ${0.4 \text{-} 0.5}$ in dense or high-multiplicity environments~\cite{ALICE:2017jyt, Andersson:1983ia, Sjostrand:2007gs}.
If a similar enhancement applies in the relatively non-equilibrium hadronic plasma following reheaton decay, the corresponding strange-hadron fraction could increase by an order of magnitude, 
\begin{align}\label{eq:Bstrange_G_enhanced}
    \Bs^{G,\text{enhanced}} \simeq \gamma_s^2 \approx 0.16 \text{-} 0.25 \ .
\end{align}
As Eq.~\eqref{eq:fS_parametrization} indicates, this directly boosts the total formation efficiency ${f_S}$, relaxing the coalescence requirement by a comparable factor.

A further enhancement may arise in asymmetric reheaton scenarios, which we discuss below in sec.~\ref{sec:asymmdecay}.
Specifically, a CP-odd reheaton coupling can produce unequal baryon and antibaryon yields with a fractional asymmetry ${\epsilon\sim 10^{-8} \text{-} 10^{-6}}$ sufficient to explain the observed baryon asymmetry for ${T_R/m_\phi\sim10^{-3} \text{-} 10^{-1}}$~\cite{Allahverdi:2010im}.
Because sexaquarks carry ${B=2}$, such asymmetric decays naturally generate a commensurate sexaquark asymmetry, linking the baryon and dark matter densities.
If reheaton decays also favor strange quarks, the combination of enhanced $\Bs$ and CP-violating $\epsilon$ can readily achieve the branching efficiency ${f_S^{\rm req} \sim 3\times 10^{-7} (m_\phi/T_R) (1860~\text{MeV}/m_S)}$ inferred from Eq.~\eqref{eq:fSreq}, matching the observed dark-matter density without fine-tuning.

Thus, while the kinematic thresholds in Fig.~\ref{fig:branching_ratios} limit $\Bs$ in the minimal model, genuine enhancements can occur through (i)~flavor-selective reheaton couplings, (ii)~strangeness-enriched hadronization, and (iii)~asymmetric reheaton decays that simultaneously generate baryon and sexaquark asymmetries.

\section{\label{sec:coalescence}Coalescence probability for compact \texorpdfstring{$uuddss$}{uuddss} formation}

At temperatures ${T \lesssim T_R < \TQCD}$, the formation of a compact sexaquark is a strongly non-equilibrium process. 
 {We parametrize the coalescence probability into two components: one from two singly-strange baryons (the $YY$ channel) and one from a doubly-strange baryon and a nucleon (the $YN$ channel).
These baryon pairs all carry the same $uuddss$ quark content as the sexaquark and contribute comparably to the overlap with the compact six-quark wavefunction~\cite{Farrar:2023wvm}.}
A convenient parametrization for the  {$YY$ channel} coalescence probability is
\begin{align}\label{eq:coal}
  \Pcoal^{(YY)} = f_Y^{\,2} \times \mathcal{P}_{YY\to S}\ ,
\end{align}
where $f_Y$ denotes the strange-baryon fraction within the relevant correlation volume, and $\mathcal{P}_{YY\to S}$ is the conditional probability that two nearby strange baryons with the correct quantum numbers combine into a compact $S$ rather than rescattering.

 {In the low-reheating scenario with ${T < \TQCD}$, the dominant channels contributing to Eq.~\eqref{eq:coal} come from the pairs of octet baryons $\Lambda\Lambda$, $\Sigma^0\Sigma^0$, and $\Sigma^+\Sigma^-$. 
We consider solely isospin-conserving reactions involving pairs of octet baryons, light mesons, and sexaquarks.}

 {In the early Universe, the plasma remains close to chemical equilibrium on timescales much shorter than the Hubble time~\cite{Kolb:1990vq}, and strangeness-changing processes are efficient at the temperatures of interest.
Thus, while separate chemical freeze out of strange and non-strange hadrons has been suggested to best fit data from heavy-ion collisions~\cite{Chatterjee:2013yga, Bugaev:2013sfa}, such effects are not expected to occur in this cosmological setting.
Strange mesons such as kaons are present in the plasma.
They rapidly equilibrate with the strange-baryon population through strong interactions.
Such meson-baryon exchanges preserve the net strangeness in the correlation volume relevant for sexaquark formation.}

We model  {the} microscopic conversion probability  {$\mathcal{P}_{YY\to S}$} as
\begin{align}\label{eq:yys}
  \mathcal{P}_{YY\to S} &\simeq \kappa_6 \times \Theta \left(p_0 - \abs*{\vec p_{\rm rel}} \right) \frac{\sigma_{YY\to S X}}{\sigma_{YY}^{\rm tot}} \nonumber \\
  &\approx \kappa_6 \left( \frac{p_0}{\LQCD} \right)^{3} \ ,
\end{align}
where ${p_0\sim 200 \text{-} 400}$~MeV is an effective coalescence momentum  {encoding the kinematic requirement that the two incoming baryons lie within a relative momentum cell of size $\sim p_0^3$,} and $\kappa_6$ gives the overlap for producing a compact spinless, flavor- and color-singlet six-quark state. 

In the second line, we used the standard coalescence scaling  {factor $\left(p_0 / \LQCD \right)^{3}$} familiar from light-nuclei formation~\cite{Scheibl:1998tk}.
 {
This geometric factor is independent of the internal color, spin, and flavor structure of the coalescing system; it applies equally to baryon-baryon, diquark-based, and six-quark assembly pathways, which we investigate further in Sec.~\ref{sec:overlap}.
Differences between these various assembly mechanisms are encoded exclusively in $\kappa_6$.}

Using ${\LQCD \simeq 300~\text{MeV}}$, the  {coalescence momentum} factor gives $ {(p_0 / \LQCD)^3 \approx} 0.3~(2.4)$ for ${p_0 = 200}$~MeV~(400~MeV).
 {The scaling in the second line of Eq.~\eqref{eq:yys} is most reliable for ${p_0 \lesssim \Lambda_{\rm QCD}}$; for $p_0$ approaching or exceeding $\Lambda_{\rm QCD}$, the phase-space factor saturates. 
We therefore treat ${p_0 \sim 400~\text{MeV}}$ as the upper boundary of validity of the scaling approximation.}

We include a second contribution to $\Pcoal$ due to a  {doubly}-strange baryon  {and nucleon} channel, $\mathcal{P}_{YN \to S}$, which obeys the same functional form, but contributes linearly in the strange-baryon fraction  {$f_Y$} rather than quadratically as in Eq.~\eqref{eq:coal},
\begin{align}
    \Pcoal^{(YN)} \simeq f_Y f_N \times \mathcal{P}_{YN \to S} \ ,
\end{align}
where ${f_N \sim 1}$ is the nucleon fraction within the correlation volume.
 {In this case, the relevant channels correspond to nucleon-cascade pairs with the required $uuddss$ quark content, $n\Xi^0$ and $p\Xi^-$.}
Both  {$YY$ and $YN$} processes operate within the same coalescence framework and differ
only by the available target density and kinematics.

 {We exclude decuplet (${J=3/2}$) baryons such as $\Sigma^\ast$ or $\Omega$, since their larger masses suppress their abundance in the plasma at ${T < \TQCD}$ and the spin coupling required to form a spin-0 sexaquark state introduces an additional suppression relative to octet baryons.}

\subsection{Overlap factor \texorpdfstring{$\kappa_6$}{kappa6}}\label{sec:overlap}

The overlap factor $\kappa_6$ quantifies the probability that six quarks produced in hadronization assemble into the specific internal configuration required for a compact, color-, spin-, and flavor singlet $uuddss$ state.
Crucially, $\kappa_6$ is not a fully normalized probability relative to all possible two-baryon states (i.e., ${P(S)/(\sum_{B,B'} P(B) \times P(B') + P(S)}$); rather, it serves as an effective projection factor in color-spin-flavor space, measuring how often six quarks align into the compact singlet configuration of the $S$, whether uncorrelated or partially clustered.
It is convenient to decompose this overlap as a product of independent alignment probabilities,
\begin{align}
    \kappa_6 \simeq P_{\rm color} \times P_{\rm spin} \times P_{\rm flavor} \times P_{\rm space} \ .
\end{align}
Two broad classes of six-quark configuration exist with this quark content: the loosely bound H-dibaryon~\cite{Jaffe:1976yi}, described as a $\Lambda\Lambda$-like molecular state, and a more compact configuration consisting of three diquarks~\cite{Shahrbaf:2022upc}. 
The probability of producing the compact state depends sensitively on the pathway by which the quarks assemble.
A generic expectation of QCD dynamics is that forming a diquark is easier than assembling a full baryon. 
Three diquarks would then assemble into a six-quark object.
We evaluate $\kappa_6$ for this possibility, but also two other production routes: the coalescence of two baryons and six independent quarks.
Formally, the baryon-baryon case corresponds to the strict limit where color-singlet hadrons form first, while the general expression allows for color recombination (e.g., via diquarks) prior to full baryon formation, permitting larger overlap values.
Together, these pathways provide a realistic range of possible overlaps.
We also assume that the energies involved are high compared to the energy multiplet splitting for SU(3)$_c$, SU(2)$_s$, and SU(3)$_f$, ensuring that the relevant internal states are populated equally. 
At lower energies, this assumption becomes conservative: if the sexaquark were the lowest-energy configuration, the effective overlap could be enhanced, leading to a larger $\kappa_6$.

Color alignment follows from SU(3)$_c$ representation counting, working from the assumption that quark colors are uncorrelated prior to any assembly.
In the baryon-baryon pathway, a color-singlet baryon~($B$) appears with probability $1/27$, since only one of the 27 color states is a singlet. 
Two such singlets automatically form a singlet, giving $P_\text{color}^{(BB\to S)} \sim (1/27)^2 \times 1 = 1/729$.
In the diquark~($D$) picture, an uncorrelated quark pair resides in the antisymmetric $\bar{\textbf{3}}_c$ representation with probability $1/3$.
If all color states are equally likely, the resulting singlet fraction from three $\bar{\textbf{3}}_c$ diquarks is therefore $1/27$, yielding the same overall factor. 
Using a sequential projection-by-pairs picture, the three diquarks combine to a singlet with a factor of $1/3$, giving instead ${P_\text{color}^{(DDD\to S)} \sim (1/3)^3 \times 1/3 = 1/81}$.
In either case, the natural scale for the color factor in the diquark route is a few percents.
Finally, when all six quarks combine directly, the total singlet weight in the full $\textbf{3}_c^{\otimes 6}$ color space gives a larger value of ${P_\text{color}^{(q^6 \to S)} \sim 5/729}$.
These three estimates together bracket the expected color-alignment probability.

Spin alignment proceeds in parallel.
For three spin-1/2 quarks, the chance of forming a spin-1/2 baryon is $1/2$. 
Two spin-1/2 baryons combine to a total-spin singlet with probability $1/4$, giving ${P_\text{spin}^{(BB \to S)} \sim (1/2)^2 \times 1/4 = 1/16}$.
In the diquark channel, an unpolarized pair of quarks forms a spin-0 scalar diquark with probability $1/4$, and combining three scalar diquarks leads to ${P_{\rm spin}^{(DDD \to S)} \sim (1 / 4)^3 \times 1 = 1 / 64}$.
Alternatively, recoupling six spin-1/2 quarks without assuming pre-formed diquarks gives a slightly higher factor ${P_{\rm spin}^{(q^6 \to S)} \sim 5 / 64}$. 
Near-threshold dynamics favoring $s$-wave, spin-0 pairs can interpolate between these two limits

Flavor introduces an additional suppression because a compact sexaquark must lie in the flavor-singlet component of SU(3)$_f$.
Scalar diquarks are antisymmetric in flavor, forming a $\bar{\textbf{3}}_f$ of SU(3)$_f$.
When three such diquarks combine, the tensor product $\bar{\textbf{3}}_f^{\otimes 3}$ contains a flavor singlet, but the singlet occupies only a small fraction of the total representation space, with the majority of states residing in octet and higher multiplets.
A representation-agnostic way to encode this limited singlet weight is to include a suppression of
\begin{align}
    P_\text{flavor} \sim 10^{-1} \text{-} 10^{-2} \ ,
\end{align}
consistent with the small singlet multiplicity obtained by explicit SU(3)$_f$ decomposition.

Finally, spatial overlap is incorporated directly through the coalescence factor $(p_0 / \LQCD)^3$, which enforces the requirement that the participating two baryons, three diquarks, or six uncorrelated quarks overlap with sufficiently small relative momentum to form an $s$-wave compact state. 
Any additional geometric suppression can be absorbed into the correlation-volume definition and into the hadronization fraction $f_Y$.
Consequently, within the coalescence framework, we take $P_{\rm space} \sim 1$.

Multiplying the individual color, spin, and flavor factors yields combined overlap estimates for each of the three assembly pathways. 
In the two-baryon pathway, ${\kappa_6^{(BB)} \simeq 9 \times (10^{-6}\text{-}10^{-5})}$. 
For the three-diquark pathway, the total color probability depends on whether one uses the rigorous SU(3)$_c$ decomposition of $\bar{\textbf{3}}_c^{\otimes 3}$ or the sequential projection picture. 
Using equiprobable colors, ${\kappa_6^{(DDD)} \simeq 6 \times (10^{-5}\text{-}10^{-4})}$, while the sequential projection gives ${\kappa_6^{(DDD)} \simeq 2 \times (10^{-5}\text{-}10^{-4})}$.
These values are naturally larger than in the baryon pathway because fewer intermediate constraints are imposed and diquark pre-clustering facilitates color-spin alignment.
In the six-quark pathways, the factor is ${\kappa_6^{(q^6)} \simeq 5 \times (10^{-5}\text{-}10^{-4})}$.
Taken together, these scenarios span a plausible envelope for the total overlap factor, ${\kappa_6 \sim 10^{-5} \text{-} 10^{-3}}$, with the precise value depending on the degree to which hadronization favors diquark clustering or direct six-quark assembly.
In practice, $\kappa_6$ should be treated as a dynamical overlap factor, rather than purely combinatorial. Near-threshold production, strong scalar-diquark attraction, and other doorway resonances can enhance $\kappa_6$; conversely, highly inelastic, decoherent production drives it to smaller values.
For phenomenological benchmarks, scanning $\kappa_6$ over the range
\begin{align}\label{eq:kappa6}
     \kappa_6 \sim 10^{-6} \text{-} 10^{-2} 
\end{align}
is sensible, keeping in mind that ${\kappa_6 \sim 10^{-2}}$ would require both strong diquark correlations and favorable kinematics.

\subsection{Encounter dynamics and reaction doorways}

The coalescence efficiency depends critically on the environment in which the reheaton decays. 
If $\phi$ decays in a hadronic plasma at ${T \lesssim \TQCD}$, the background contains a thermal population of nucleons with density ${n_B = \eta n_\gamma}$. 
This baryon abundance is not the one inherited from reheaton decay; rather, it reflects the equilibrium value sustained by rapid hadronic processes such as $\gamma\gamma \leftrightarrow B \bar{B}$ and multi-meson reactions.
Above the temperature at which baryon-antibaryon annihilation freezes out, these reactions equilibrate any baryon produced in reheaton decay with the thermal bath.
Once freeze out occurs, the baryon density falls out of equilibrium and becomes sensitive to the injection history.
Thus, depending on whether $\phi$ decays before or after baryon freeze-out, the reheaton decay products either merge into the thermal bath population or generate an out-of-equilibrium abundance that subsequently cools and free-streams.

A strange baryon $Y$ produced promptly in the reheaton decay cascade can coalesce into an $S$ if it undergoes an encounter with another baryon before it free-streams or decays.
The probability of such a re-encounter is 
\begin{align}
    P_\text{enc} \sim n_\text{targ} \sigma_\text{had} L_\text{eff} \ ,
\end{align}
where $\sigma_\text{had} \sim 20$-$40$~mb is a strong hadronic cross section and $L_\text{eff} \sim \mathcal{O}(1~\text{fm})$ is the interaction length.
The target density $n_\text{targ}$ may correspond to the ambient nucleon density $n_N$, or in the case of a low-energy hadronic jet, to a local microburst of correlated hadrons.
Such microburst can raise $n_\text{targ}$ by many orders of magnitude over the homogeneous background within the small volume relevant for coalescence, greatly enhancing the chance of forming the compact six-quark singlet.

The timescale hierarchy is crucial.
The weak decay lifetimes of strange octet baryons are ${\tau_Y \simeq 10^{-10}~\text{s}}$, whereas coalescence, being governed by QCD, occurs on a timescale of order ${1/\LQCD \sim 10^{-23}~\text{s}}$.
Therefore, as long as the interaction rate ${\Gamma_Y \sim n_\text{targ} \ev{\sigma v}}$ exceeds the weak decay rate, strange baryons interact multiple times before decaying.
This condition is readily satisfied for temperatures ${T \gtrsim 10~\text{MeV}}$, even with only ambient nucleons as targets.
Below this threshold, strange baryons would mostly decay before a second strong interaction, and only nucleon-nucleon encounters would remain available for coalescence.

Given this kinetic window for rescattering, we now examine the two primary reaction mechanisms (or ``doorways'') that facilitate coalescence into $S$: the double-hyperon~($YY$) channel and the single-hyperon~($YN$) channel.

\subsubsection{\texorpdfstring{$Y + Y$}{Y+Y} (double-hyperon) doorway}

If hadronization generates a pair of hyperons (e.g., $\Lambda\Lambda$, $\Xi N$) with significant phase-space overlap, they may undergo a direct fusion or resonant $\mathcal{R}$ transition to a sexaquark:
\begin{align}
    Y(p_1)+Y(p_2) \  (\to \mathcal{R})  \to S + X \ ,
\end{align}
where ${Y \in \lbrace \Lambda, \Sigma, \Xi, \Omega \rbrace}$.
Here, $X$ denotes soft mesons needed to conserve quantum numbers and carry off binding energy. 
This mechanism can realize ${\mathcal{P}_{YY\to S} \sim 10^{-4} \text{-} 10^{-2}}$ provided the pair has small relative momentum (${\abs*{\vec p_{\rm rel}} \lesssim p_0}$) and the overlap factor $\kappa_6$ is not too suppressed.
This channel benefits strongly from low-multiplicity, few-GeV jets where strangeness is locally conserved.

\subsubsection{\texorpdfstring{$Y+N$}{Y+N} (single-hyperon) doorway}

This channel utilizes the ambient baryon asymmetry as the second participant, producing one slow hyperon $Y$ per decay:
\begin{align}
    Y + N \to S + X \ ,
\end{align}
where $X$ represents light particles which do not carry baryon number, such as mesons (typically $\pi$ or $K$) and photons.
If the $S$ is deeply bound in the flavor-singlet $I=0$, $J=0$ channel~\cite{Jaffe:1976yi, Inoue:2010es, Farrar:2017eqq}, the reaction can be exothermic.
The per-encounter conversion probability can be ${\mathcal{P}_{YN \to S} \sim 10^{-3}}$ in narrow near-threshold kinematics. 
This doorway avoids the penalty of producing a pair of strange baryons in the same micro-jet and helps with antibaryon management by consuming ambient nucleons.

\subsection{Increasing the coalescence probability}

Coalescence is most efficient when the relative momentum between the coalescing hadrons is small and the system evolves under near-nonrelativistic kinematics.
In this regime, the wavefunction overlap between two baryons in the ${I=0}$, ${J=0}$ channel and a compact $uuddss$ state becomes maximal.
Both lattice-QCD calculations and phenomenological models indicate that near-threshold $Y N$ and $Y Y$ systems can exhibit attractive interactions sufficiently strong to transiently form correlated di-hyperon states or molecular precursors to the $S$~\cite{Jaffe:1976yi, Inoue:2010es, Farrar:2017eqq, Scheibl:1998tk}.
Such near-threshold rescattering provides an effective ``doorway'' for forming a compact flavor-singlet bound state before the hyperons weakly decay.
Quantitatively, coalescence probabilities $\mathcal{P}_{YY\to S}$ as large as ${10^{-4} \text{-} 10^{-2}}$ can be achieved for relative momenta ${|\vec p_{\mathrm{rel}}|\lesssim p_0\simeq200\text{-}400~\mathrm{MeV}}$ and for overlap factors $\kappa_6$ near their upper theoretical limits.
These conditions may be realized in low-multiplicity, few-GeV hadronic environments or in dense post-hadronization plasmas, where repeated $Y$-baryon scattering occurs on timescales ${\tau_{\rm QCD}\ll\tau_Y}$, ensuring multiple opportunities for $S$ formation before weak decay.

\subsection{Range of \texorpdfstring{$f_S$}{fS} from reheaton decay}

\begin{figure}[ttt]
    \centering
    \includegraphics[width=\linewidth]{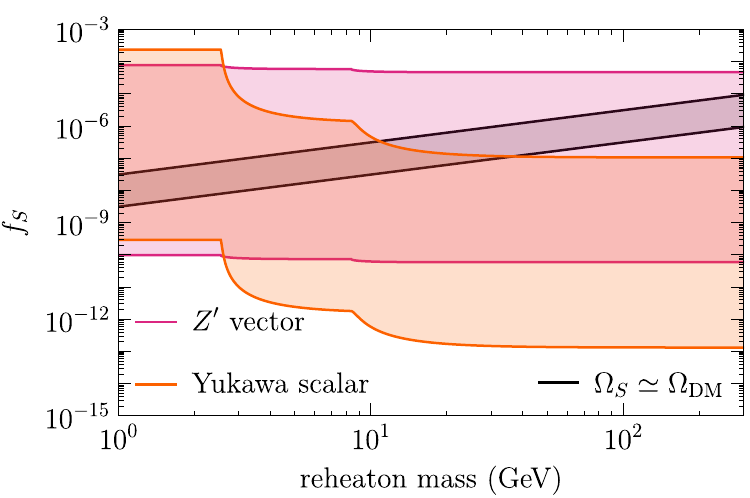}
    \caption{Fraction $f_S$ of reheaton decay producing sexaquarks as a function of the reheaton mass. The grey band indicates the region where the full dark matter abundance in sexaquarks can be obtained for the reheating temperature range considered throughout this work, ${10 \leq T_R \leq 100~\text{MeV}}$, with smaller reheating temperatures corresponding to the top of the band. The colored bands indicate the range obtained for two scenarios; a scalar Yukawa and a flavor-universal vector reheaton. The width of each region is determined by the uncertainty on the parameters that control $f_S$. 
    }
    \label{fig:fS_summary}
\end{figure}

Here, we combine the ranges obtained in Sec.~\ref{sec:Bstrange} for $\Bs$ and Sec.~\ref{sec:coalescence} for $\Pcoal$.
We assess the resulting values of $f_S$, defined in Eq.~\eqref{eq:fSreq}, the effective probability for producing stable sexaquarks from reheaton decay, for (pseudo)scalar, gluophilic, and (axial-)vector reheatons across the relevant mass thresholds.
Figure~\ref{fig:fS_summary} summarizes the numbers found from this section for scalar Yukawa and flavor-universal vector reheatons, with qualitatively similar results for pseudovector Yukawa and flavor-universal axial-vector reheatons, respectively.

First, we combine the overlap factor and dynamics.
Allowing the full ranges for spin-color-flavor overlap factor $\kappa_6$ from Eq.~\eqref{eq:kappa6} and effective coalescence momentum ${p_0 \in [200, 400]~\text{MeV}}$, we obtain
\begin{align}
    \mathcal{P}_{YY \to S} \sim 10^{-7} \text{-} 10^{-2} \ .
\end{align}
Combining with the strange baryon fraction $f_Y$ in the range $f_Y \sim 10^{-3} \text{-} 10^{-2}$, we find the coalescence probability to be
\begin{align}
    \Pcoal^{(YY)} \sim 10^{-13} \text{-} 10^{-6} \ ,
\end{align}
with the upper end corresponding to optimistic assumptions about color-spin projection and low relative momentum in jet-like microbursts. 
Similarly, the coalescence probability for the single-hyperon channel gives
\begin{align}
    \Pcoal^{(YN)} \sim 10^{-10} \text{-} 10^{-4} \ .
\end{align}
We now combine $\Pcoal^{(YN)}$ with the strange-branching fractions $\Bs$ characteristic of the different reheaton scenarios to determine the resulting range for $f_S$.

\begin{itemize}
    \item \textbf{(Pseudo)scalar Yukawa reheaton:} 
    The range for $\Bs^Y$ is given in Eq.~\eqref{eq:Bs_Yukawa}.
    Near but below the charm threshold, (pseudo)scalar reheatons decay almost exclusively into strange quarks, $\Bs^Y \simeq 1$. 
    In this regime, the full range of $f_S$ is directly inherited from $\Pcoal$. 
    For larger reheaton masses, the charm and then bottom quark channels open up, strongly reducing the strange branching fraction. This yields
    \begin{align}\label{eq:fS_Yukawa}
        f_S^Y = \begin{cases} 
            10^{-10} \text{-} 10^{-4} & 2 m_s \lesssim m_\phi \lesssim 2 m_c \\
            10^{-12} \text{-} 10^{-6} & 2 m_c \lesssim m_\phi \lesssim 2 m_b \\
            10^{-13} \text{-} 10^{-7} & 2m_b \lesssim m_\phi
        \end{cases} \hspace{-0.5cm}
    \end{align}

    \item \textbf{Gluophilic reheaton:}
    For gluophilic reheatons, the strange fraction arises from hadronization of gluon jets and is less sensitive to heavy-flavor thresholds. The conservative range for $\Bs^G$ is given in Eq.~\eqref{eq:Bs_gluon}. We find
    \begin{align}\label{eq:fS_gluon}
        f_S^G = 
        \begin{cases}
            10^{-11} \text{-} 10^{-5} & \text{conservative} \\
            10^{-11} \text{-} 10^{-4} & \text{enhanced }s\text{ production}
        \end{cases} \hspace{-0.5cm}
    \end{align}

    \item \textbf{Flavor-universal (axial-)vector reheaton:}
    For vectors and axial vectors with flavor-universal couplings, the strange branching fraction $\Bs^U$, Eq.~\eqref{eq:Bs_universal}, is set primarily by quark multiplicities and phase space. 
    The resulting range is
    \begin{align}\label{eq:fS_universal}
        f_S^U = 10^{-10} \text{-} 10^{-4} \ .
    \end{align}
    Compared to scalar reheatons at the same mass, vector and axial-vector scenarios populate systematically larger values of $f_S$, owing to their less suppressed strange-quark branching fractions.
    Flavor-nonuniversal reheaton couplings that suppress charm and bottom relative to strange quarks could maintain $\Bs^U \simeq \mathcal{O}(1)$, even above heavy-flavor thresholds.
\end{itemize}

Across all reheaton realizations, the dominant sources of variation in $f_S$ are the strange branching fraction $\Bs$ and the poorly known coalescence parameters ($\kappa_6, p_0, f_Y$). 
Scalar and pseudoscalar reheatons below the charm threshold maximize $\Bs$, but suffer strong suppression once heavier flavors open up, while flavor-universal vectors and gluophilic reheatons maintain sizable $\Bs$ values over a broad mass range, shown in the more substantial overlap between the grey and purple regions in Fig.~\ref{fig:fS_summary}. 
Taken together, the resulting values of $f_S$ span several orders of magnitude, due to the model dependence and significant theoretical uncertainties of $\Bs$ and $\Pcoal$, and overlap with the values required to reproduce the observed dark matter abundance, see Fig.~\ref{fig:fS_summary}.

\subsection{Comparison with previous studies}

An alternate estimate of the efficiency with which strange quarks are trapped in sexaquark compared to other strange hadrons is given by~\cite{Farrar:2018hac}
\begin{align}
    \kappa_s(m_S, T) = \frac{1}{1 + (r_{\Lambda,\Lambda} + r_{\Lambda,\Sigma} + 2 r_{\Sigma,\Sigma} + 2 r_{N,\Xi})} \ ,
\end{align}
where $r_{i,j}$ are products of phase space factors of baryons $i$ and $j$ divided by the phase space factor of the sexaquark, and given by ${r_{i,j}\equiv \exp(-[m_i + m_j - m_S]/T)}$. 
This expression is based on a restrictive equilibrium treatment; considering only octet baryons, no mesons or decuplet states, no spin-weight corrections accounting for the fact that octet baryons are spin-$\frac{1}{2}$ particles while the sexaquark has zero spin. 
Its physical meaning is somewhat ambiguous once strangeness-violating weak decays become rapid compared to Hubble. 
Its value, for a sexaquark mass of $m_S \sim 1860$~MeV, is smaller for larger reheating temperatures, giving $\kappa_s \gtrsim 0.7$, which appears overly optimistic.

The coalescence of three diquarks into a compact sexaquark, for varying sexaquark radii, was investigated using the PACIAE model in the context of proton-proton collisions in a TeV collider~\cite{She:2025dqx}. However, the kinematics differ greatly from those occurring between the QCD crossover and BBN, such that it is not possible to compare our estimate to their results.

In a separate work, the possibility of forming bound states of helium-3 and antiprotons, then annihilating into light mesons and sexaquarks~\cite{Doser:2023gls}.
The kinematics are closer to those we consider, but the scheme differs by the presence of antimatter.

Theoretical estimates for the projection of two octet baryons with same overall quark content as the sexaquark, such as $\Lambda\Lambda$, suggest a group-theoretical upper bound of ${\kappa_6 \leq 1/40}$, arising from the color-singlet projection ${P_\text{color} \sim 1/5}$ and spin-flavor overlap ${P_\text{spin} \times P_\text{flavor} \sim 1/8}$~\cite{Farrar:2022mih, Farrar:2023wvm}.
The color factor is enhanced relative to our estimate and arises from the potential internal organization of colors, which might not need to be identical to two color-singlet baryons.
Beyond the $\textbf{1}_c \otimes \textbf{1}_c$ singlet we considered, the $\textbf{8}_c \otimes \textbf{8}_c$ would couple to form a singlet, while both ``baryons'' would be color octets.

\section{\label{sec:freeze-in}Freeze-in coalescence during reheating}

In addition to the direct production of sexaquarks via prompt reheaton decays, we discuss here a parallel production pathway: sexaquark formation via \emph{thermal} hadronic
encounters during the matter-dominated reheating phase with ${T_R<\TQCD}$. 
The Hubble rate during this epoch is
\begin{equation}
    H(T) \simeq A \frac{T^{4}}{M_{\rm Pl} T_R^{2}} \ ,
    \qquad
    A \equiv \sqrt{\frac{\pi^{2} g_\ast}{90}} \ ,
\label{eq:H_MD}
\end{equation}
where $g_\ast$ is the effective number of relativistic degrees of freedom contributing to energy density. 
The entropy density is given by ${s(T)=\frac{2\pi^2}{45}g_s T^3}$, where $g_s$ counts the entropic degrees of freedom.
The comoving yield from rare hadronic reactions ${i j\to SX}$ is
\begin{equation}
\begin{aligned}
    Y_S^{\rm(FI)} \equiv \frac{n_S}{s}
    &\simeq
    \int_{T_R}^{T_{\max}} \dd{T}
    \frac{n_i n_j \ev{\sigma v}_{ij\to S X}}{s(T) H(T) T}\\[4pt]
    &=
    \int_{T_R}^{T_{\max}} \dd{T}
    \frac{\zeta T^{6} e^{-2m_{\rm eff}/T}}{s(T) H(T) T} \ ,
\end{aligned}
\label{eq:Yint}
\end{equation}
where we parametrize the Boltzmann-suppressed interaction rate below $\TQCD$ and hadronic overlap by
\begin{equation}
    n_i n_j \ev{\sigma v}_{ij\to SX}
    \equiv \zeta T^{6} e^{-2m_{\rm eff}/T} \ ,
\end{equation}
where $\zeta$ is an effective coupling constant with dimensions of $\text{GeV}^{-2}$ (encoding multiplicities and overlap probabilities), and $m_{\rm eff}$ is the effective mass threshold for the relevant channel,
\begin{equation}
\begin{aligned}
    m_{\rm eff}
    &\sim
    \begin{cases}
        \tfrac{1}{2}(m_Y+m_N) & (Y N~{\rm doorway}) \ ,\\[2pt]
        m_Y & (Y Y~{\rm doorway}) \ .
    \end{cases}
\end{aligned}
\label{eq:param}
\end{equation}

The effective coupling $\zeta$ encodes strange-hadron availability, six-quark overlap, and $S$-wave coalescence into a single factor:  
\[
\zeta = \mathcal{C}_{\rm had}\times \left[f_Y^{\,2} \kappa_6 \big(p_0/\Lambda_{\rm QCD}\big)^{3}\right] \ .
\]
To estimate the hadronic normalization factor $\mathcal{C}_{\rm had}$ one starts from a hadronic scale cross section and folds in a near-threshold suppression: using ${1~{\rm GeV}^{-2}=0.389~{\rm mb}}$, a typical hadronic size ${\sigma_{\rm had}\sim 20 \text{-}40~{\rm mb}}$ corresponds to ${50 \text{-} 100~{\rm GeV}^{-2}}$; multiplying by a conservative ${\chi_{\rm kin}\sim 10^{-4} \text{-} 10^{-6}}$ (accounting for the small phase-space window that actually coalesces) gives
\[
\mathcal{C}_{\rm had} \sim 10^{-2}\text{-} 10^{-4}~{\rm GeV}^{-2} \ .
\]
With ${f_Y \sim 10^{-3}}$, ${(p_0/\Lambda_{\rm QCD})^{3} \sim 0.03 \text{-} 0.1}$, ${\kappa_6 \sim 10^{-4}}$, and  $\mathcal{C}_{\rm had}$ estimated as above, one obtains ${\zeta \sim 10^{-29} \text{-} 10^{-27}~{\rm GeV}^{-2}}$, while aggressively optimistic choices can push to ${\zeta \sim 10^{-22}~{\rm GeV}^{-2}}$.


Carrying out the integral with Eq.~\eqref{eq:H_MD} gives a closed-form solution dominated by the highest plasma temperature $T_{\max}$:
\begin{equation}
\begin{aligned}
    Y_S^{\rm(FI)}
    &\simeq
    \mathcal{N}(T_R,m_{\rm eff}) \times e^{-2m_{\rm eff}/T_{\max}} \ ,
\end{aligned}
\label{eq:YFI-closed}
\end{equation}
with the prefactor
\begin{equation}
    \mathcal{N}(T_R,m_{\rm eff})
    \equiv
    \frac{\zeta M_{\rm Pl} T_R^2}
    {\big(\tfrac{2\pi^2}{45}\big) g_s A} 
    \frac{1}{2m_{\rm eff}}\ .
\label{eq:Nprefactor}
\end{equation}

For ${g_* \simeq g_s \simeq 10.75}$ the prefactor simplifies to 
${\mathcal{N} \simeq 0.098 \zeta M_{\rm Pl} T_R^2/m_{\rm eff}}$. Considering a benchmark with reheating temperature ${T_R =100~{\rm MeV}}$, maximum temperature ${T_{\max}= 150~{\rm MeV}}$, and effective mass threshold ${m_{\rm eff} \sim (1.0 \text{-}1.1)~\text{GeV}}$, we find the Boltzmann factor to be strongly suppressed, ${e^{-2m_{\rm eff}/T_{\max}} \lesssim 2\times 10^{-6}}$.
This suppression is generic: whenever the maximum temperature of the thermal bath remains well below the relevant mass threshold, freeze-in production is exponentially Boltzmann suppressed, independently of the microscopic realization of the interaction~\cite{Cosme:2023xpa, Cosme:2024ndc}.
The prefactor evaluates to
\begin{equation}
    \mathcal{N} 
    \approx 1.2\times 10^{16} \zeta \ ,
\label{eq:Nnumerics}
\end{equation}
resulting in a final freeze-in yield of
\begin{equation}
    Y_S^{\rm(FI)}
     \lesssim (2\text{-} 3)\times 10^{10} \zeta
    \ .
\label{eq:YFI-num}
\end{equation}
To reproduce the observed dark matter abundance, which corresponds to a yield of ${Y_{\rm DM} \simeq 2.3 \times 10^{-10} (1.86~{\rm GeV}/m_S)}$
, the effective coupling parameter would need to satisfy ${\zeta \sim 10^{-20}~\text{GeV}^{-2}}$.
This requirement is physically implausible: as discussed above, even with optimistic assumptions for strange hadron densities and cross sections, one expects ${\zeta \lesssim 10^{-22}~\text{GeV}^{-2}}$ (and more realistically ${\zeta \lesssim 10^{-28}~\text{GeV}^{-2}}$ for standard thermal inputs).
Consequently, achieving the full dark matter density via freeze-in would require significantly higher temperatures (${T_{\max} \gtrsim \mathcal{O}({\rm GeV})}$) or implausibly large strange-hadron multiplicities to overcome the exponential suppression.
We thus regard freeze-in coalescence
as parametrically \emph{subdominant} in the low-reheat regime of interest.

\section{\label{sec:antiparticles}Antibaryon and antisexaquark depletion}

In cosmological histories where (anti)baryons and (anti)sexaquarks are generated in chemical equilibrium with the photon bath during the QCD crossover from the quark-gluon plasma epoch to the hadron epoch, these species carry a small baryon chemical potential~\cite{Gross:2018ivp, Kolb:2018bxv, Moore:2024mot}. 
Subsequent rapid interactions with the thermal bath allow to efficiently deplete the antibaryon number, leaving the observed baryon-to-photon ratio ${\eta \equiv n_b / n_\gamma}$~\cite{PDG:2024}.
Here, the hadronic states are generated at a later time. 
It is essential that any residual antibaryon or antisexaquark population be driven to negligible levels before the onset of BBN.
We characterize this requirement for antibaryons and antisexaquarks, and we determine the constraints on the presence of a non-negligible antisexaquark population.

\subsection{Antibaryon depletion}

 {As detailed in Section~\ref{sec:asymmetry}, the reheaton must decay through baryon-number-violating~(BNV) interactions that simultaneously violate CP, or there must be additional mechanisms that do at a later time. 
This generates an asymmetry between baryons and antibaryons, establishing a non-zero baryon chemical potential ${\mu_b \not = 0}$ in the post-reheating hadronic plasma.}
Given this asymmetry, we evaluate whether there is enough time to deplete the antibaryon yield  { through annihilation reactions}.

Any mechanism that produces baryon-antibaryon pairs during or just after reheating must ensure that the residual antibaryon abundance is driven to a negligible level before BBN.
The thermally-averaged annihilation cross section for $p\bar{p}$ or $n\bar{n}$ is ${\ev{\sigma v}_{B\bar{B}} \sim 30 \text{-}60}$~mb in the relevant kinematic regime.
Heavier baryonic states could also be produced and their decay rate into the lightest baryon species is much shorter than the timescale from the end of reheating to the onset of BBN, such that their production does not raise any issue.
Even up to BBN, the annihilation rate ${\Gamma = n_B \ev{\sigma v}_{B \bar{B}}}$ remains much larger than Hubble, ensuring that the antibaryons annihilate away even if the reheating temperature is not far above $\TBBN$.

\subsection{Antisexaquark annihilation bounds}\label{sec:asymmbounds}

 {The presence of a baryon number violation from the reheaton decay or a similar mechanism establishes a chemical potential $\mu_b$ in the early hadronic plasma.
This chemical imbalance satisfies ${\mu_S = 2\mu_b}$ due to the sexaquark's ${B = 2}$ quantum number.
Here, we evaluate whether the residual antisexaquarks are sufficiently depleted before BBN.}

The annihilation cross section for the sexaquark is expected to be $s$-wave, due to the scalar nature of the particle, and at least geometric, with
\begin{align}\label{eq:sigma_geom}
    \sigma_\text{geom} \sim \pi r_S^2 \sim 10^{-28} \text{-} 10^{-27}~\text{cm}^2 \ ,
\end{align}
where the sexaquark radius is $r_S \sim 0.1$-$0.4$~fm~\cite{Farrar:2017eqq}.
The corresponding thermally-averaged annihilation cross section $\ev{\sigma v}$ is approximately $\ev{\sigma v} \sim \sigma_\text{geom} v$, which at recombination (${v_\text{rec} \sim 10^{-5}c}$), results in $\ev{\sigma v}_\text{rec} \gtrsim 9 \times 10^{-25}~\text{cm}^3 \, \text{s}^{-1}$. 
Comparatively, cosmic microwave background~(CMB) energy injection limits obtained from the \textit{Planck} satellite require~\cite{Planck:2018vyg, Slatyer:2015jla}
\begin{align}
    p_\text{ann} \equiv f_\text{eff} \frac{\ev{\sigma v}_\text{rec}}{m_S} \lesssim 3 \times 10^{-28}~\text{cm}^3 \, \text{s}^{-1} \, \text{GeV}^{-1} \ ,
\end{align}
where $f_\text{eff}$ is the effective deposition efficiency due to dark matter annihilation.
For hadronic final states, as would be expected from $S\bar{S}$ annihilation, ${f_\text{eff} \sim 0.3 \text{-}0.6}$, implying ${\ev{\sigma v}_\text{rec} \lesssim 2 \times 10^{-27}~\text{cm}^3 \, \text{s}^{-1}}$, which exceeds the hadronic expectation.
\textit{Fermi-LAT} $\gamma$-ray limits from dwarf spheroidal galaxies and annihilation constraints into neutral pions which then decay to pairs of photons, yield comparable, or even stronger constraints~\cite{Fermi-LAT:2015att, Moore:2024mot}.

Given that a symmetric sexaquark population is generally excluded in the absence of a large suppression of the annihilation cross section below the hadronic scale, we investigate the allowed antisexaquark fraction in light of current bounds. 
We define ${r = n_{\bar{S}}/n_S \leq 1}$ as the residual fraction of antisexaquarks~\cite{Graesser:2011wi}. 
For a total sexaquark mass density ${\rho_S = m_S (n_S + n_{\bar{S}})}$, the annihilation term ${R \equiv \ev{\sigma v} n_S n_{\bar{S}}}$ becomes
\begin{align}
    R = \frac{\rho_S^2}{m_S^2} \ev{\sigma v}_\text{eff} \ ,
\end{align}
where we defined the effective thermally-averaged annihilation cross section ${\ev{\sigma v}_\text{eff} \equiv \ev{\sigma v} r/(1 + r)^2}$. 
The constraints from \textit{Planck} on CMB energy injection imply {$r \lesssim 10^{-6} $-$ 10^{-5}$}~\cite{Planck:2018vyg, Slatyer:2015jla}, while \textit{Fermi-LAT} dwarf galaxy constraints typically requires ${r \lesssim 10^{-7} \text{-} 10^{-6}}$~\cite{Fermi-LAT:2015att}, where the exact value depends on $f_\text{eff}$ and halo kinematics.
A complementary, largely astrophysics-independent stringent limit comes from the extragalactic diffuse $\gamma$-ray background, giving similar constraints to \textit{Fermi-LAT}, and tightening slightly when halo clustering (structure boosts) are included~\cite{Chen2025_EGB, Abdo:2010EGB, Ackermann:2015IGRB, Strong:2004EGB, Ajello:2015EGBorigin}.

Collectively, these constraints demonstrate that any viable asymmetric scenario demands an extremely small residual antisexaquark fraction, at or below the part-per-million level.
In this regime, $\ev{\sigma v}_\text{eff}$ is suppressed by at least six orders of magnitude relative to the geometric annihilation rate, pushing conventional indirect-detection signals beyond current sensitivity.
The resulting phenomenology is therefore dominated by the asymmetric component, with only highly suppressed annihilation signals.

Reheaton decays would inject both sexaquarks and antisexaquarks into the plasma. 
The viability of this scenario depends critically on the reheat temperature $T_R$ relative to the $S\bar{S}$ annihilation freeze-out temperature $\Tfo$.
For ${T_R < \Tfo}$, the annihilation rate is negligible, and any antisexaquark produced in the decay cannot be eliminated.
Without a mechanism to suppress their production, these relic antisexaquarks would survive and give rise to a strong signature in CMB and dwarf galaxy observables.
For instance, the thermally-averaged annihilation cross section $\ev{\sigma v}_\text{eff} \simeq 10^{-26}~\text{cm}^3 \, \text{s}^{-1}$ would turn off at 90~MeV. 
To avoid the problematic existence of antisexaquarks, either ${T_R > \Tfo}$ is needed, such that antisexaquarks have time to annihilate away, or the reheaton decay must be fundamentally asymmetric by producing an excess of sexaquarks over antisexaquarks.

This requirement implies the existence of new CP-violating phases in the reheaton sector, as the quark mixing matrix phase is insufficient to generate large asymmetries.
Furthermore, at low reheating temperature electroweak sphalerons are exponentially suppressed, preventing the reprocessing of a potential lepton asymmetry into a baryon asymmetry.
Therefore, both the baryon and sexaquark asymmetries must, most likely, be generated directly at ${T \lesssim T_R}$.
Mechanisms for baryogenesis at low reheating temperatures have been extensively studied~\cite{Davidson:2000dw}
. We review a few possible such mechanisms in the following section.


\section{\label{sec:asymmetry}Asymmetry generation: baryons and sexaquarks}

As a result of the discussion above, it is clear that any viable sexaquark dark matter scenario requires not only achieving the correct relic abundance, but also ensuring that the residual antisexaquark population is sufficiently suppressed to evade the discussed, stringent indirect detection constraints. 
 {This requires BNV interactions in the reheaton sector, also necessary for addressing the matter-antimatter asymmetry.}
In the standard thermal freeze-out picture with high reheating temperature ${T_R \gg \TQCD}$, the sexaquark inherits an asymmetry from the baryon chemical potential established during some necessary, additional high-temperature baryogenesis scenario.
$S\bar{S}$ annihilation at ${T \sim \Tfo \sim 90~\text{MeV}}$ is not sufficient to deplete the symmetric component to negligible levels.
Efficient annihilations must continue until lower temperatures, ${T \sim 40~\text{MeV}}$, to deplete the symmetric component, at the expanse of underproducing dark matter in sexaquarks.
In contrast, non-thermal production via late reheaton decay raises a new question: how is the $S$-$\bar{S}$ asymmetry generated when sexaquarks are produced directly from decays at ${T_R < \TQCD}$?

 {Here, we discuss three concrete mechanisms through which the reheaton can generate both the baryon asymmetry and the necessary sexaquark asymmetry simultaneously in the low-reheating scenario.
They all}
tie naturally to the broader challenge of low-temperature baryogenesis, and to the appealing possibility that non-thermal sexaquark dark matter be explicitly tied to baryogenesis.

\subsection{Asymmetric reheaton decay}\label{sec:asymmdecay}

The most direct mechanism is for the reheaton $\phi$ itself to decay asymmetrically into matter versus antimatter through CP-violating couplings  {in addition to the Yukawa or gauge couplings that produce strange quarks, discussed in Section~\ref{sec:Bstrange}.
The Yukawa couplings in Eq.~\eqref{eq:pseudoscalar} and vector couplings in Eq.~\eqref{eq:vector} conserve baryon number individually}. 
If the reheaton carries CP-violating phases beyond the Cabibbo-Kobayashi-Maskawa~(CKM) matrix, its decays can satisfy the Sakharov conditions~\cite{Sakharov:1967dj}  {and generate both baryon and sexaquark asymmetries}:
\begin{itemize}
    \item \textbf{Baryon number violation:} The reheaton  {couples to BNV operators, enabling} decays into hadronic final states that  {violate} net baryon number.
    \item \textbf{C and CP violation:}  {The BNV couplings introduce new phases that allow interference between tree- and loop-level decay amplitudes.}
    The decay rates satisfy ${\Gamma(\phi \to B + X) \neq \Gamma(\phi \to \bar{B} + X)}$, where $B$ represents baryonic matter (including sexaquarks with ${B=2}$) and $X$ denotes  {light} accompanying SM particles.
    \item \textbf{Out-of-equilibrium conditions:} The reheaton decays occur while ${\Gamma_\phi \lesssim H}$, ensuring the asymmetry is not washed out by inverse processes.
\end{itemize}
This scenario would naturally generate both the observed baryon asymmetry ${\eta \sim 6 \times 10^{-10}}$ and a corresponding sexaquark asymmetry in a unified framework.
The required CP violation must come from new physics in the reheaton sector, as the CKM phase is insufficient to generate large asymmetries~\cite{Gavela:1993ts, Huet:1994jb}, and electroweak~(EW) sphalerons are exponentially suppressed at ${T \ll T_{\rm EW}}$.

For example, consider a CP-odd reheaton coupling (such as the pseudoscalar coupling $y_{qP}$ in Eq.~\eqref{eq:pseudoscalar}) or a complex scalar with CP-violating interference between tree-level and loop-induced decay amplitudes.
Such couplings can yield decay asymmetries of
\begin{equation}
    \epsilon \equiv \frac{\Gamma(\phi \to B) - \Gamma(\phi \to \bar{B})}{\Gamma(\phi \to B) + \Gamma(\phi \to \bar{B})} \sim \mathcal{O}(10^{-8} \text{-} 10^{-6}) \ ,
\end{equation}
sufficient to simultaneously generate $\eta$ and the required sexaquark-to-baryon ratio. 
 {The coefficient $\epsilon$ encodes CP violation in the BNV sector.
Importantly, this CP violation is independent of the mechanism producing strangeness.}
For a late-decaying scalar that reheats the Universe, the resulting baryon yield is ${Y_B \simeq \epsilon (3T_R/4m_\phi)}$ akin to Eq.~\eqref{eq:yield_sexaquark}~\cite{Allahverdi:2010im}, so ${\epsilon \sim 10^{-8} \text{-} 10^{-6}}$ reproduces the observed $\eta$ for the ratio ${T_R/m_\phi \sim 10^{-3} \text{-}10^{-1}}$ and branching fraction  {of reheatons into the lightest hadrons (protons and neutrons) as ${f_B \sim 1}$.
Note that the reheaton must carry baryon number or couple to BNV operators to enable this mechanism.}

 {Two regimes are relevant, depending on whether $T_R$ lies above or below the 
$S\bar{S}$ annihilation freeze-out temperature ${T_{\rm fo} \simeq 90}$~MeV. For ${T_R \gtrsim T_{\rm fo}}$, the symmetric component 
of the sexaquark yield is efficiently depleted by $S\bar{S}$ annihilation, leaving 
a net dark matter yield ${Y_S \simeq \epsilon \times f_S (3T_R/4m_\phi)}$. Reproducing 
${\Omega_S = \Omega_{\rm DM}}$ in this case requires ${\epsilon \times f_S}$ to satisfy 
Eq.~\eqref{eq:fSreq}, implying ${f_S \sim f_S^{\rm req}/\epsilon \sim 10^{7}\text{-} 10^{9}}$, which is 
not achievable within our coalescence framework. For ${T_R \lesssim T_{\rm fo}}$, 
which covers most of our benchmark range, $S\bar{S}$ annihilation is already frozen 
at the time of reheating, and any antisexaquarks produced cannot be subsequently 
depleted. In this regime the reheaton decay must be \emph{intrinsically} asymmetric, 
directly producing a highly suppressed antisexaquark fraction ${r \equiv 
n_{\bar{S}}/n_S \lesssim 10^{-6}}$ from the outset. The surviving dark matter 
abundance is then set by the total yield ${Y_S \simeq f_S(3T_R/4m_\phi)}$ of 
Eq.~\eqref{eq:yield_sexaquark}, with the asymmetry $\epsilon$ simultaneously generating the observed baryon 
asymmetry $\eta$ and ensuring the required suppression of $n_{\bar{S}}$. This is 
the physically consistent realization of the asymmetric reheaton scenario within our 
benchmark parameter space.}

\subsection{Low-temperature baryogenesis scenarios}

If the reheaton does not directly generate the asymmetry  {(baryon-number violation and CP violation reside in a separate sector)}, it can still facilitate baryogenesis through secondary processes in the post-reheating hadronic plasma.
Several such mechanisms for generating baryon asymmetry at temperatures $T \lesssim \TQCD$ have been explored in the literature:

\subsubsection{Affleck-Dine baryogenesis at low reheat}

In supersymmetric extensions, flat directions in the scalar potential can carry a large baryon or lepton number.
If these fields decay after reheating is complete, their CP-violating decays can generate the observed asymmetry even at ${T_R \sim 10 \text{-}100}$~MeV~\cite{Affleck:1984fy, Dine:1995kz}.
Sexaquarks, carrying ${B=2}$, would naturally inherit a corresponding asymmetry from the same BNV decays.

\subsubsection{Spontaneous baryogenesis}

If a scalar field with derivative couplings to the baryon current evolves during or after reheating, its time-dependent background can generate a chemical potential for baryons~\cite{Cohen:1987vi, Cohen:1988kt}.
This mechanism operates out of equilibrium and can function at arbitrarily low temperatures, provided the relevant interactions remain active.

\subsubsection{Leptogenesis with late decays}

Heavy right-handed neutrinos or other exotic fermions decaying at $T \sim T_R$ can generate a lepton asymmetry, which is partially converted to a baryon asymmetry via sphaleron processes at ${T > T_{\rm EW}}$ or through lower-temperature mechanisms~\cite{Fukugita:1986hr}.
However, for $T_R \ll T_{\rm EW}$, electroweak sphalerons are inactive, so direct baryon-number violation in the reheaton sector is required.

\subsection{Implications for antisexaquark abundance}

Regardless of the specific mechanism, successful asymmetry generation must satisfy two requirements:
\begin{enumerate}
    \item \textbf{Sufficient asymmetry:}
    The ratio of antisexaquarks to sexaquarks must satisfy ${r \equiv n_{\bar{S}}/n_S \lesssim 10^{-6}}$ to evade CMB energy injection and \textit{Fermi-LAT} dwarf galaxy constraints (see the discussion in Sec.~\ref{sec:asymmbounds} above).
    \item \textbf{Thermal viability:}
    If ${T_R < T_{\rm fo} \sim 90~\text{MeV}}$ (the $S\bar{S}$ annihilation freeze-out temperature), any produced antisexaquarks cannot be depleted by annihilations. In this regime, the reheaton decay (or subsequent meson decays) \textit{must} be intrinsically asymmetric, as there is no post-production annihilation epoch to remove the symmetric component.
\end{enumerate}
For reheating temperatures ${T_R \gtrsim 50 \text{-}100~\text{MeV}}$, one can in principle produce sexaquarks and antisexaquarks symmetrically and rely on residual $S\bar{S}$ annihilations to deplete the antisexaquark population, provided the annihilation cross section is sufficiently large.
However, for lower reheating temperatures or to minimize fine-tuning of annihilation rates, generating an intrinsic asymmetry via the mechanisms outlined above is essential.

Low-temperature baryogenesis frameworks 
thus offer a natural and phenomenologically viable route to producing asymmetric sexaquark dark matter in the non-thermal regime.
These mechanisms tie the sexaquark asymmetry directly to the observed baryon asymmetry, providing a unified origin for both components of the cosmic matter budget.


\section{\label{sec:summary}Summary}

The viability of non-thermal sexaquark production mechanisms described in this work is governed by three ingredients.
The first is the branching fraction of reheaton decays into strange quarks, $\Bs$, which controls the rate at which strange baryons capable of forming a $uuddss$ state are injected in the low-temperature plasma.
The second is the coalescence probability $\Pcoal$, encoding the likelihood that the strange baryons emerging from the reheaton decay assemble into a tightly bound sexaquark rather than ordinary hyperons.
The third is the cosmological efficiency with which the resulting sexaquarks survive the entropy injection associated with reheaton decay, encoded in the ratio $T_R/m_\phi$.

The resulting dark matter abundance depends linearly on the production efficiency $f_S$ and the dilution factor $T_R/m_\phi$, as detailed in Eq.~\eqref{eq:yield_sexaquark}.
This relationship incorporates the full entropy production from the decay: any sexaquarks (or other relics) produced \textit{before} the Universe becomes $\phi$-dominated are diluted by the same entropy injection that lowers all pre-existing yields, whereas sexaquarks produced in the decay itself scale linearly with the reheaton yield.
Quantitatively, for fixed microphysics ($f_S$ and $m_S$), the combination $T_R/m_\phi$ therefore measures the net dilution: lowering $T_R/m_\phi$ corresponds to increasing the entropy release per reheaton particle and thus suppressing the final $S$ yield.
As shown in Fig.~\ref{fig:omegaS}, we find that non-thermal production of sexaquarks can account for all of dark matter for ${f_S \sim 10^{-6}}$.

The effective production efficiency $f_S$, introduced in Eq.~\eqref{eq:fSreq}, is the product of the strange branching fraction $\Bs$ and the coalescence probability $\Pcoal$.
We evaluated $\Bs$ for three main scenarios: a (pseudo)scalar Yukawa interaction with quarks, a gluon-coupled (``gluophilic'') scalar, and a flavor-universal (axial-)vector particle, as detailed in Eqs.~(\ref{eq:Bs_Yukawa}), (\ref{eq:Bs_gluon}), (\ref{eq:Bs_universal}) and shown in Fig.~\ref{fig:branching_ratios}, finding that for a Yukawa interaction, at worst ${\Bs \gtrsim 4 \times 10^{-4}}$ without any fine-tuning, with larger values obtained below the threshold to produce bottom quarks, as well as in the other interaction channels.
We then determined the two dominant contributions to $\Pcoal$: from the coalescence of two strange baryons (labeled $YY$ channel) and one strange baryon with a nucleon (labeled $YN$ channel).
For generic strong-scale coalescence kinematics with ${p_0\sim 0.2 \text{-} 0.4~\text{GeV}}$, the microscopic coalescence probability is 
${\Pcoal \sim \kappa_6 (p_0/\LQCD)^3  \sim 10^{-1} \text{-}1}$, depending on the overlap factor $\kappa_6$.

Since both $\Bs$ and $\Pcoal$ are model-dependent, combining them leads to a broad range of possible values for $f_S$, as illustrated through the width of the colored bands in Fig.~\ref{fig:fS_summary}, and presented in Eqs.~(\ref{eq:fS_Yukawa})--(\ref{eq:fS_universal}) for a (pseudo)scalar Yukawa, gluophilic, and (axial-)vector reheaton, respectively.
For representative ${T_R/m_\phi \sim 10^{-3} \text{-} 10^{-2}}$, some of these values falls short of the $f_S^{\rm req}$ values needed to obtain ${\Omega_S=\Omega_{\rm DM}}$ by one to three orders of magnitude.


We pointed out several physically motivated effects which can raise $f_S$ toward the required level:
\begin{enumerate}
    \item Strangeness-rich hadronization can enhance $\Bs$ by up to an order of magnitude.
    If the effective strangeness-suppression factor $\gamma_s$ increases from the nominal ${0.2 \text{-} 0.3}$ to ${\sim 0.4 \text{-}0.5}$ in the dense, non-equilibrium hadronic plasma following reheaton decay, as suggested by high-multiplicity collider data, then ${\Bs \simeq \gamma_s^2 \approx 0.16 \text{-} 0.25}$ and $f_S$ can reach $ 10^{-4}$ (see Eq.~\eqref{eq:fS_gluon}).
    
    \item The non-thermal production history circumvents the exponential Boltzmann suppression that plagues the thermal relic scenario.
    In the freeze-in regime, where the sexaquark never attains chemical equilibrium with the baryon bath, the comoving yield scales linearly with the total production rate rather than as $\exp(-m_S/T)$.
    This ``incomplete equilibration'' condition, ${\Gamma_{YY\to S} \lesssim H}$ at ${T \sim 100~\text{MeV}}$, is naturally realized for coalescence cross sections below the geometric hadronic scale, and the associated freeze-in contribution -- though typically subdominant to direct reheaton decay -- provides a robust floor for the sexaquark abundance, without being exponentially suppressed. 

    \item An asymmetric reheaton decay can simultaneously generate both the baryon and sexaquark asymmetries.
    A CP-odd coupling yields a decay asymmetry ${\epsilon \sim 10^{-8} \text{-} 10^{-6}}$ sufficient to reproduce ${\eta \simeq 6\times 10^{-10}}$ for ${T_R/m_\phi\sim 10^{-3} \text{-} 10^{-1}}$, while sexaquarks with ${B=2}$ inherit a commensurate dark-sector asymmetry ${Y_S/Y_B\simeq m_p/(2m_S) \approx 0.25}$.
    Because both baryons and sexaquarks are produced in the same reheaton decay and diluted by the same entropy injection, their ratio $Y_S/Y_B$ is insensitive to the absolute normalization of $T_R/m_\phi$.
    Entropy production in reheaton decay thus suppresses {\it all} pre-existing relics but leaves the baryon-sexaquark ratio set by $f_S$ and $\epsilon$ essentially unchanged.
\end{enumerate}

Taken together, these considerations indicate that the non-thermal production of sexaquark dark matter is a quantitatively viable possibility, provided at least one of the following conditions holds:
(i)~enhanced strangeness production with ${\Bs \gtrsim 0.1}$, 
(ii)~coalescence probabilities ${\Pcoal \gtrsim 0.1}$ in dense hadronic environments, 
(iii)~freeze-in dynamics preventing exponential suppression, or 
(iv)~an asymmetric reheaton decay producing linked baryon and sexaquark asymmetries.
While each factor alone is insufficient, their combination -- especially (ii)+(iv) with a favorable $T_R/m_\phi$ -- naturally yields the correct relic abundance for a stable ${m_S\simeq 1860~\text{MeV}}$ state.
The reheaton-driven non-thermal scenario therefore remains a compelling and testable pathway for sexaquark dark matter.

Finally, while this work has focused on the cosmological production of sexaquarks, it is important to emphasize that a stable ${m_S \simeq 1860 \text{–} 1890}$~MeV state constituting dark matter today is subject to strong direct-detection constraints~\cite{Moore:2024mot}.
Current underground searches exclude most of the parameter space for strongly interacting dark matter in this mass range,  {ruling out scenarios in which a $uuddss$ state behaves as a nucleon-sized hadron with unsuppressed hadronic interactions, i.e. a scattering cross section of order $\sigma_\text{geom}$ (see Eq.~\eqref{eq:sigma_geom}). 
A compact, flavor-singlet, scalar sexaquark can naturally evade direct-detection limits: one-pion exchange is suppressed due to isospin, axial couplings vanish, and elastic scattering proceeds predominantly through short-distance gluonic operators that are strongly suppressed at the low momentum transfers relevant for underground experiments~\cite{Farrar:2022mih}.
In such scenarios, the effective nucleon-sexaquark cross section can lie many orders of magnitude below current limits while remaining consistent with atmospheric and terrestrial stopping constraints~\cite{Scherrer:1985zt}.
Consequently, the reheaton-driven production mechanism studied here is compatible with present-day constraints only if the sexaquark is effectively neutral under strong nuclear interactions at low energies at low energies.
Direct detection excludes only a restricted class of strongly interacting realizations}, leading to ${\sigma_{Sn} \lesssim 10^{-42}~\text{cm}^2}$~\cite{Cirelli:2024ssz}{, and does not provide a model-independent ruling out of compact sexaquark dark matter}.
The reheaton-driven scenario studied here is therefore compatible with present-day constraints only if the sexaquark is effectively ``hadronically dark’’ at low energies, with suppressed couplings to ordinary nuclei despite its intrinsic QCD nature.

\section*{Acknowledgments}

M.M. is supported by the U.S. Department of Energy, Office of Science, Office of High Energy Physics of U.S. Department of Energy under grant Contract Number DE-SC0012567, and by the Simons Foundation (Grant Number~929255). S.P. is supported in part by the U.S. Department of Energy, Office of Science, Office of High Energy Physics under grant Contract Number DE-SC010107.


\bibliography{biblio}

\end{document}